\documentclass[aps,preprint,pra,preprint]{revtex4}

\usepackage{bm}
\usepackage{amsmath}
\usepackage{amssymb}
\usepackage{graphicx}
\usepackage{epsfig}
\usepackage{epstopdf}

\begin{document}

\def\be{\begin{equation}}
\def\en#1{\label{#1}\end{equation}}
\def\d{\dagger}
\def\bar#1{\overline #1}
\newcommand{\per}{\mathrm{per}}
\newcommand{\eqb}{\begin{eqnarray}}
\newcommand{\eqe}{\end{eqnarray}}
\newcommand{\rd}{\mathrm{d}}
\newcommand{\bx}{{\bf x}}
\newcommand{\bk}{{\bf k}}
\newcommand{\bq}{{\bf q}}
\newcommand{\br}{{\bf r}}
\newcommand{\bQ}{{\bf Q}}
\newcommand{\rX}{{\rm X}}
\newcommand{\vare}{\varepsilon }
\newcommand{\pd}{\partial}

 \title{Asymptotic evaluation of  bosonic  probability amplitudes in linear unitary networks in the case of large number of bosons    }

\author{V. S. Shchesnovich}

\address{Centro de Ci\^encias Naturais e Humanas, Universidade Federal do
ABC, Santo Andr\'e,  SP, 09210-170 Brazil \\ 
E-mail: valery@ufabc.edu.br}

\begin{abstract}
An   asymptotic analytical approach is proposed for    bosonic probability amplitudes in unitary linear networks, such as the  optical multiport devices for photons. The asymptotic approach applies for large number of bosons $N\gg M$ in the $M$-mode  network, where $M$ is finite. The probability amplitudes of  $N$ bosons unitarily transformed from the input  modes   to the output  modes of a unitary network  are approximated by a multidimensional integral    with the integrand  containing  a large parameter ($N$) in the exponent. The integral representation allows an asymptotic estimate of   bosonic probability amplitudes up to a multiplicative error of order $1/N$ by the saddle point method. The estimate depends on  solution of the scaling problem for the $M\times M$-dimensional  unitary  network matrix: to find the left and right diagonal matrices which scale the unitary matrix to a matrix which has specified   row and column sums (equal, respectively, to the distributions of bosons in the input and output modes). The scaled matrices give the saddle points of the integral.  For simple saddle points, an explicit formula giving the asymptotic  estimate of  bosonic probability amplitudes  is derived. Performance of the  approximation and the scaling  of the relative  error with $N$ are  studied  for two-mode network (the beam-splitter),  where the saddle-points are  roots of a quadratic and an exact  analytical formula for the probability amplitudes is available, and  for  three-mode network  (the tritter).

\end{abstract}

\maketitle

\section{Introduction}

Linear optical networks  play an important role in the quantum manifestations of light. Indeed, the  well-known Hong-Ou-Mandel (HOM) dip \cite{HOM}   (see also Refs.~\cite{CST,LM}) is a direct  manifestation of the quantum indistinguishability  of  photons. Recently \cite{LB} a generalization of the HOM effect and  difference in  behavior of bosons and fermions was  theoretically studied    in the general setting of   Bell multiport beam splitters  (see also, for instance,  Refs.~\cite{Mult1,Mult2}). Many   important results on  quantum interference phenomena in  multiport devices for bosons and fermions were recently discovered. There is  a zero transmission law \cite{ZTL} in    Bell multiports  due to   symmetry of the network matrix.    Moreover, a generalization of the suppression laws and many-particle interferences beyond the  boson and fermion statistics in   unitary linear networks are  found \cite{MPI}.   Recently  the experimental advances have allowed to verify   the HOM effect  and  the zero transmission law   with three photons on a tritter \cite {HOM2zero}. There are   other important  experimental advances in  quantum interference experiments with indistinguishable particles in linear multiport devices, for instance, the  recent three-photon quantum interference experiment  on an integrated eight-mode optical device \cite{MPH}. Moreover, the experimental efforts are now also directed at building the so-called  boson sampler \cite{E1,E2,E3,E4} (see also below). Finally,  a linear bosonic network   is the central  part of an ingenious  proposal of quantum computation based on  linear  
optics \cite{KLM}.

It is known that the outcome  probability amplitudes in unitary  bosonic networks (e.g. in  optical multiports)  are expressed through  matrix permanents, somewhat  similar to the fermonic amplitudes which are  given by  matrix (a.k.a. Slater)   determinants.  Considering $N$ non-interacting  bosons in an unitary  network of $M$ input and output modes, one has to compute  the matrix permanents of  the $N\times N$-dimensional  complex  matrices composed of repeated  rows and columns of the network  matrix to find the bosonic transition amplitudes in the network. The  permanent of a matrix  
 \cite{Minc}  can be obtained by taking the well-known Laplace  expansion   formula for  matrix determinant  and  setting  to ``$+$''  the signatures of \textit{all} permutations in the summation.    
 
 The relation of  bosonic amplitudes to matrix permanents   was  explored   a long time ago \cite{C} in connection with  the quantum fields theory.      
Recently  it has attracted  a lot of renewed attention. One reason is that the similarity between fermions and bosons does not go along if one tries to compute  matrix permanent: unlike it is for  matrix determinant,  computation of the  permanent of an arbitrary matrix is  $\#P$-complete, which is the result of a classic paper in the computational complexity theory \cite{Valiant}.  This means that no algorithm polynomial in matrix size  can  compute  matrix permanents. The  fastest  known algorithm for computing the permanent of an arbitrary (complex) $n\times n$-dimensional  matrix is based on   Ryser's formula \cite{Ryser} and  requires $\mathcal{O}(n^22^{n})$ flop operations.     An interesting    ``physical" proof of the matrix permanent complexity result, based on  the  linear optical computing proposal \cite{KLM},  was recently discovered \cite{A1}.    Moreover, even     the problem of approximating the  permanent of an arbitrary complex matrix  to a polynomial multiplicative error was also shown \cite{A1} to be   a $\#P$-hard problem.\footnote{There is an  important exception of  permanents of matrices with   positive elements,  which  can be effectively approximated \cite{JSV}, but such permanents  are not connected to  the  \textit{quantum} transition amplitudes in linear networks.} Complexity of the matrix permanent was analyzed   in  the context of  quantum mechanics in Ref.~\cite{TT}, where  a method based on  quantum measurement was proposed to directly  measure the matrix permanent in a   bosonic quantum  multiport device. It was shown that the  permanent  of an arbitrary matrix can be expressed as a  quantum observable. However, the catch of the method lies in an  exponential  number of  necessary  measurements, since the variance of the   observable giving matrix permanent  is exponential in matrix size \cite{TT} (for comparison it was shown that  matrix  determinant  can be  found  in just  a single  measurement).

Deep connection between the  complexity of bosonic networks and that of  matrix permanents was used  in the recent proposal of a new model of   quantum computer  with noninteracting identical bosons,  which,  though not being  an universal quantum computer, nevertheless  can perform computations considered to be hard on a classical computer \cite{A1,AA}.  Such new quantum computer  was compared  to classical  Galton's board where, instead of  classical balls and a single entry point,  identical bosons are launched into    different  modes of  a  linear network. A very crucial difference, however, is  that in the quantum network case  the bosonic probabilities themselves are not known beforehand, since they cannot be  effectively  computed  on a classical computer (for a sufficiently  large network).   One has to run  the actual sampling experiments to find  them. In the technical part, the proposal depends on the  hardness to approximate the  matrix permanent of a large  $N\times N$-dimensional submatrix of  an arbitrary  unitary $M\times M$-dimensional  matrix      and some numerically tested    conjectures.  This proposal has generated  experimental efforts to build the necessary   bosonic network \cite{E1,E2,E3,E4} (see also the related experiments of Refs.~\cite{HOM2zero,MPH}).

It is believed that the hardness of  computing the permanent of an arbitrary    matrix  is related to the matrix rank. Computation of  the matrix permanent of an arbitrary (complex) $n\times n$-dimensional  matrix of rank $R_n$    would apparently require at least on the order  of $ n^{R_n}$ flop operations  on a classical computer (see, for instance, Ref.~\cite{Gurvits}; moreover,  this  estimate  for  a matrix with repeated columns or rows is derived  in Appendix D).   In terms of bosonic networks, different limits are possible  in this respect, which can be  roughly   divided by the relation between the number of bosons $N$ (i.e. matrix size)  and the number of modes $M$  in the network (i.e. the maximum  rank). For instance, it is  shown \cite{AA} that even an approximate computation of the matrix permanents  describing   linear  bosonic networks  is $\# P$-hard  in the limit of   $N\to \infty$ and  $M \gg N^2$. The latter inequality  is related  to the ``boson birthday bound'', i.e.  the  probability of two bosons to land into the same mode  is negligible in this limit (and there is an  experimental   confirmation \cite{BosBirthExp}).   

 The computational complexity of  matrix permanent is also related to the fact that the permanent, in contrast to the determinant, takes different values on equivalent matrices, that  is $\mathrm{per}(U)\ne \mathrm{per}(VUV^{-1})$, for arbitrary invertible $V$. Nevertheless, similar to the determinant, the permanent may be evaluated exactly due to some symmetry of the   matrix (i.e. in the  case of a quantum network, due to a destructive interference of   bosonic  transition  amplitudes).  For instance, some exact results are known for  Schur's matrices \cite{GL}.  Recently,    suppression laws for bosons and fermions  were derived  for  Bell multiports \cite{ZTL,MPI}  by using  such a symmetry.  These laws apply to  arbitrarily large Bell multiport devices and thus are very important conceptually due to their generality.

But  even for Bell multiports there is no analytical approach capable to  give   probabilities of individual events when the latter  are greater than zero (however, probabilities of averaged output events, such as    the probability to find a specific number of bosons  in one output port or a specific number of occupied output ports was in fact approximated by an analytical formula based on the classical consideration \cite{ZTL}).
Such an  analytical approach  is  especially   lacking in the  limit of a large matrix   size, when the computational complexity makes  numerical evaluation   practically hard.  One would not expect the existence of an asymptotic  analytical approximation in the limit $N\to\infty$ and $M\ge N$, since   it would   contradict the $\#P$-hardness of numerical approximation of the permanent in this limit (indeed,  one could then run a numerical scheme mimicking the analytical approach). An asymptotic  analytical approximation  could only exist  for $M\ll N$. Such an approach is developed  below by reducing the summation   of the Laplace expansion of  matrix  permanent  to a multidimensional  integral of the saddle-point type in the limit $N\to \infty$ and $M$   finite. \textit{The very existence of  an asymptotic approximation is important and stems  from the fact that it is not dependent on the  matrix size $N$ (i.e. the number of bosons in the network), but only  on the solution of some $2M-1$ bilinear equations (the matrix scaling problem) giving the saddle points of the integral.}

 Practical usage of the proposed approach  crucially depends on effective solution of two other problems of different type. One is the unitary matrix scaling problem, i.e. for a given unitary matrix $U$ to find all diagonal scaling matrices $X$ and $Y$  with the complex-valued  elements such that the product matrix $XUY$ has given row and column sums (these matrices give  the saddle points of the integral). The other  problem is to derive a  formula for multidimensional  saddle point method  with the multiple, i.e. coalescing,   saddle points. Only the   case of simple saddle points has a general solution so far \cite{FB} and a special case of  two coalescing  saddle points on the real axis  is treated \cite{FP}.  However, the coalescing saddle points represent  exceptional cases rather than the general case.  For simple saddle points,  which is the general case, an explicit formula for bosonic probability amplitudes is derived below.  As  we use  the well-known Stirling formula  for the factorial  in one of the steps, the approach  is restricted  to the case when   there is no input or output mode without at least one   boson in it (though this latter limitation can be lifted, for simplicity sake it is not pursued in this work).  The asymptotic  formula for  bosonic transition amplitudes is tested  on  two-mode network (i.e. the beam splitter) where it shows a very good accuracy within the above  described limitations. Three-mode network (i.e. the tritter) is also considered.

The rest of the text is organized as follows. In section \ref{sec2} an integral approximation of bosonic probability amplitudes in unitary   networks is derived. The integral is evaluated by the saddle point method and an explicit formula is derived when the saddle points are simple. Some details of the calculations are relegated to \ref{AppB} and \ref{AppC}.  For completeness, a short derivation of bosonic probability amplitudes as  matrix permanents is given in \ref{AppA}. Comparison with the classical particles on a network is also discussed. In section \ref{sec3} the general formula is tested on two-mode network (the beam splitter) where the sources of  error are identified and discussed. Moreover,  three-mode network (the tritter) is also considered. Section \ref{sec4} contains summary of the results. The  computational complexity of the permanent of a matrix with repeated rows and/or columns, i.e. for which the asymptotic approximation is developed,  is  considered in detail in Appendix D.

\section{The saddle point method for probability amplitudes in linear bosonic  networks}
\label{sec2}
\vskip 2.5cm
\begin{figure}
\begin{center}
\includegraphics[width=0.6\textwidth]{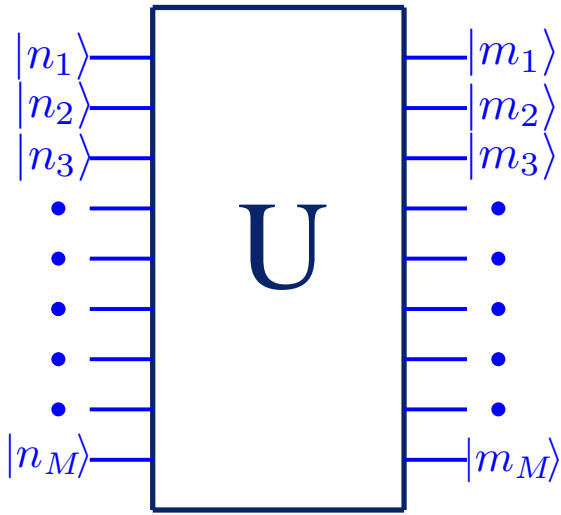}
 \caption{  A schematic (black-box) depiction  of    bosonic network described by an unitary matrix $U$, where $U_{kl} = \langle g_l|f_k\rangle$.  The  Fock states in the input modes $|f_1\rangle,\ldots,|f_M\rangle$  and those in the output modes $|g_1\rangle,\ldots,|g_M\rangle$ are indicated by 
 $|n_1\rangle,\ldots,|n_M\rangle$ and    $|m_1\rangle,\ldots,|m_M\rangle$, respectively (the numbers $n_1,...,n_M$ and $m_1,...,m_M$  are also the numbers of repetitions   of  rows and columns of the network matrix $U$ in the matrix $U[n_1,...,n_M|m_1,...,m_M]$ in Eq. (\ref{E1})).}
 \end{center}
\end{figure}

We consider the transition probability amplitudes of $N$ bosons between the    input ($|f_1\rangle,\ldots,|f_M\rangle$) and   output  ($|g_1\rangle,\ldots,|g_M\rangle$) modes of a $M$-mode  unitary network, as in  Fig.~1,  which are given by (see, for instance, Refs.~\cite{TT,Scheel})
\be
{}_g\langle m_{1},\ldots,m_{M}|n_{1},\ldots,n_{M}\rangle_f
= \frac{\mathrm{per}(U[n_1,\ldots,n_M|m_1,\ldots, m_M])}{\sqrt{\prod_{k=1}^Mn_k!m_k!}}.
\en{E1}
Here  matrix $U[n_1,\ldots,n_M|m_1,\ldots,m_M]$ consists of repeated rows and columns of the network matrix $U_{kl} = \langle g_l|f_k\rangle$ (the order being insignificant),  with  $n_k$ duplicates of the $k$th row and  $m_l$ of the $l$th column,  satisfying $\sum_{k=1}^M n_k = \sum_{k=1}^M m_k = N$ (for more details, see \ref{AppA}). Recall that the permanent of a $N\times N$-dimensional matrix $A$ is given by summation over all possible permutations $\tau$ in the product of  $N$ different matrix elements \cite{Minc}, i.e. 
\be
\mathrm{per}(A) = \sum_{\tau} A_{1\tau(1)}\cdot\ldots\cdot A_{N\tau(N)}.
\en{per}

To derive an asymptotic approximation for the   bosonic amplitude given  in Eq. (\ref{E1}) we  regroup of the summation in Eq. (\ref{per}) in such a way that the nature of the matrix $U[n_1,\ldots,n_M|m_1,\ldots,m_M]$  as a matrix with repeated rows and columns is  used. This regrouping leads also to an interesting interpretation of the   permanent of $U[n_1,\ldots,n_M|m_1,\ldots,m_M]$ as an average over a lattice of contingency tables, see Fig.~2,  with probabilities given by a bi-multivariate generalization of the  hypergeometric distribution, known as the Fisher-Yates distribution in mathematical statistics \cite{Fisher,Yates} (see also a number of reviews Refs.~\cite{Good,DE,DG} and the references therein).  This representation is discussed below in detail.

\subsection{Representation of bosonic permanent as an average over a lattice of contingency tables}

By  successive application of the  Laplace expansion for  permanents \cite{Minc} one can express the boson probability amplitude (\ref{E1}) as an  average over the lattice of $M\times M$-dimensional   matrices $S_{kl}$ ($S_{kl}\in \{0,1,2,3\ldots\}$) with  given row and column sums, equal here to the distribution of bosons in the input and output modes: $\sum_{l=1}^M S_{kl} = n_k$ and $\sum_{k=1}^M S_{kl} = m_l$. Indeed,  the first application of the Laplace expansion consists of dividing the matrix $U[n_1,\ldots,n_M|m_1,\ldots,m_M]$ into two parts, and the permanent into a sum over the  products of  partial permanents, one  involving $n_1$ first rows (all equal)  and the other involving the rest of  matrix rows. The summation runs over  all possible partitions  of the column indices between  the two permanents,  such that one permanent contains $n_1$ column indices ($w_1,\ldots,w_{n_1}$, with $S_{11}$ of them being  equal to $1$, $S_{12}$ being equal to $2$, etc), while  the other permanent contains  $N-n_1$ column indices ($w_{n_1+1},\ldots,w_{N}$).  Thus,   application of the Laplace  expansion  as above described gives (with $(w_1,\ldots,w_N) $ being a  permutation of $(\underbrace{1,\ldots,1}_{m_1},\underbrace{2,\dots,2}_{m_2}, \ldots, \underbrace{M,\ldots,M}_{m_M} )$)
\begin{eqnarray}
&&\per(U[n_1,\ldots,n_M|m_1,\ldots,m_M])=\sum_{w_1,\ldots,w_{N}}\per(U[n_1|S_{11},\ldots,S_{1M}])
\nonumber\\
&&\times\per(U[n_2,\ldots,n_M|m_1-S_{11},\ldots,m_M-S_{1M}])
\nonumber\\
&&= n_1!\sum_{S_{11},\ldots,S_{1M}}\delta_{\sum S_{1l},n_1}\prod_{l=1}^M\frac{m_l!}{S_{1l}!(m_l-S_{1l})!}U_{1l}^{S_{1l}}
\nonumber\\
&&\times\per(U[n_2,\ldots,n_M|m_1-S_{11},\ldots,m_M-S_{1M}]),
\label{E2}\end{eqnarray}
where we have used that permanent of the first submatrix is equal to $n_1! \prod_{l=1}^MU_{1l}^{S_{1l}}$ and, due to the permutational invariance of matrix permanent,  the r.h.s. contains functions of numbers of  repeated  columns and not the  column   indices themselves. Repeating the above procedure, by taking out successively each set of repeated rows, we obtain\footnote{Eq. (\ref{E3}) could also  be   deduced from Eq. (\ref{E1}) and the   formula for  bosonic probability amplitude derived  in Ref.~\cite{Urias} by application of  Wick's theorem; note that using the matrix permanent and the  Laplace expansion   is an impressive  shortcut to that  involved derivation.}
\begin{eqnarray}
&&\per(U[n_1,\ldots,n_M|m_1,\ldots,m_M])=\left[\prod_{k=1}^Mn_k!m_k!\right]
\nonumber\\
&&\times\sum_{S_{kl}\ge0}\left[\prod_{k=1}^M\delta_{\sum_{l=1}^M S_{kl},n_k}\right]\left[\prod_{l=1}^M\delta_{\sum_{k=1}^M S_{kl},m_l}\right]\prod_{k,l=1}^M\frac{ U_{kl}^{S_{kl}} }{ S_{kl}! }.
\label{E3}\end{eqnarray}

Eq. (\ref{E3})   has  an interesting statistical interpretation.\footnote{Besides physical interpretation as  the sum over  quantum probability amplitudes of  all possible transitions through the network, i.e. a variant of R.~Feynman's   path integral formula.}  Indeed, the ratio of  factorials which appears on the r.h.s. in Eq. (\ref{E3}), divided by $N!$, is know as   the Fisher-Yates distribution \cite{Fisher,Yates} (see also the reviews Refs.~\cite{Good,DE,DG}).  It appears in applied mathematical statistics, namely in Fisher's exact test of independence of two properties, and   uses the so-called contingency tables (here $S_{kl}$).  The Fisher-Yates distribution gives  the conditional probability of getting a matrix $S_{kl}$ (the contingency table)  of  the joint frequencies of two statistically independent properties,  given the  row and column sums are equal to their marginal frequencies  (the margins, in our case $n_1,\ldots,n_M$ and $m_1,\ldots,m_M$, respectively).  Since the two properties are independent, a simple exercise in combinatorics (see also  footnote ``d" ) leads to  the following probability formula of the Fisher-Yates distribution (using  a shortcut notation $\{v_i\}$ for a set of indexed  variables: $v_1, v_2,\ldots$)
\be
\mathcal{ P}(\{S_{kl}\}|\{n_k,m_l\}) = \frac{ \left[\prod_{k=1}^M n_k!\right] \left[\prod_{l=1}^M m_l!\right] }{ N!\prod_{k,l=1}^MS_{kl}! }.
\en{E4}
Note that the delta symbols in Eq. (\ref{E3}) restrict the summation to  matrices $S_{kl}$ with given margins and  precisely under these constraints the probabilities of Eq. (\ref{E4}) sum to 1.   The matrix permanent of Eq. (\ref{E3}) is thus  multiplied by $N!$ the value of the characteristic function $\chi(\{\lambda_{kl}\}|\{n_k,m_l\}) \equiv \langle \exp\{\sum_{k,l=1}^M\lambda_{kl}S_{kl}\}\rangle$
of the Fisher-Yates distribution at the parameters $\lambda_{kl}$ equal to  logarithms of  the elements  of $U$:
\begin{eqnarray}
&&\per(U[n_1,\ldots,n_M|m_1,\ldots,m_M]) = N!\left\langle \prod_{k,l=1}^MU_{kl}^{S_{kl}}\right\rangle
\nonumber\\
&&\qquad\qquad=N!\chi(\{\mathrm{ln}(U_{kl})\}|\{n_k,m_l\}).
\label{E5}\end{eqnarray}
Eq. ~(\ref{E5}) has the following   physical interpretation. The bosonic transition amplitude in a quantum network, between the input  
$|n_1,\ldots,n_M\rangle_f$ and  output   $|m_1,\ldots,m_M\rangle_g$ Fock states, is an average  of  products of  amplitudes of  ``elementary processes", as depicted in Fig.~2, corresponding to  contingency table $S_{kl}$ (i.e. a matrix satisfying   the constraints $\sum_{l=1}^MS_{kl}= n_k$ and $\sum_{k=1}^MS_{kl}=m_l$),  \textit{ assuming the mutual   statistical  independence  of   distribution of bosons in the input and output modes.}\footnote{There are  $C_{in} = \frac{N!}{\prod_{k=1}^Mn_k!}$ ways to distribute bosons over the input modes, $C_{out} = \frac{N!}{\prod_{l=1}^Mm_l!}$  ways to distribute bosons over the output modes, and, as the two distributions are  independent,  in total $ C_{in}C_{out}$ ways to distribute bosons over the input and output modes. For a table $S_{kl}$ we select  $C_{S}  = \frac{N!}{\prod_{k,l=1}^M S_{kl}!}$ combinations from these distributions, thus $P(\{S_{kl}\}|\{n_k,m_l\}) = \frac{C_{S}}{C_{in}C_{out}}$, i.e. Eq.~(\ref{E4}).}

\begin{figure}
\vskip 3.5cm
\begin{center}
\includegraphics[width=0.6\textwidth]{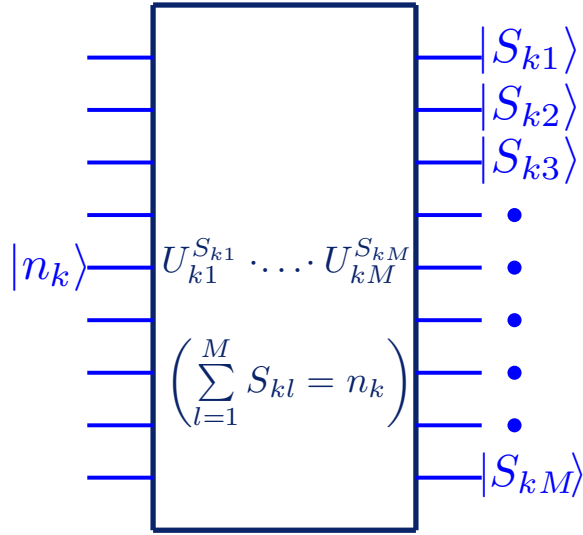}
 \caption{ ``Elementary process" $|0,\ldots,n_k,\ldots,0\rangle_f\to|S_{k1},\ldots,S_{kM}\rangle_g$ with the amplitude $\prod_{l=1}^M U^{S_{kl}}_{kl}$ (the product of  such elementary amplitudes for $k=1,\ldots,M$ gives the transition  amplitude associated with  the contingency table $S_{kl}$ in Eq. (\ref{E5})).   Here the  Fock state in the $k$th input mode $|f_k\rangle$ is  $|n_k\rangle$    and the Fock states  in the output modes $|g_1\rangle,\ldots,|g_M\rangle$  are  
 $|S_{k1}\rangle,\ldots,|S_{kM}\rangle$.}
 \end{center}
\end{figure}

Eq.  (\ref{E5}), however, does not seem to be of any help for  numerical evaluation  of matrix permanents, since the number of  contingency tables scales exponentially  with $N$ (precisely, their number scales exponentially with the margins for a fixed table size and, in fact, the problem of counting  the contingency tables is $\#P$-hard, see, for instance Refs.~\cite{DG,Bend,Barv,GM,Barv2}). 

On the other hand, computation of the permanent of a matrix with repeated rows and/or columns \textit{can be effectively carried out} by using the  available in this case  reductions in  Ryser's algorithm (see, for instance,  Ref.~\cite{Thesis}). Indeed, it is shown in Appendix D that     a modified Ryser  algorithm requiring  just $\mathcal{O}(N^{M+1})$ flops is available  in this case (this algorithm was used for obtaining Fig. 7 of section \ref{tritt}   below).

\subsection{Approximating  the bosonic probability amplitude   by a multidimensional  integral}

On the other hand, the fact that  the number of  contingency tables in Eq. (\ref{E5}) scales exponentially  with $N$ is an indication on possibility of an  asymptotic approach. Indeed,  since the lattice of contingency tables $S_{kl}$ is exponential in $N$, if we divide $S_{kl}$ by $N$,  the resulting matrix $p$,  $p_{kl}\equiv S_{kl}/N$,  will belong as $N\to \infty$ to a \textit{  dense lattice } in the    continuous convex set   of matrices with real-valued entries and  given row and column sums. Therefore,  a sum over such a lattice can be replaced by a multidimensional integral with   $p_{kl}$ as the integration variables. Moreover,  a large parameter $N$ would appear   in  the exponent of the integrand, thus  allowing for an asymptotic evaluation of the permanent.   This is the approach pursued in the following.

First of all, we need to approximate  the  Fisher-Yates distribution  (\ref{E4})  by a manageable smooth  function of  the integration variables $p_{kl}$. Using an approximate formula for the multinomial coefficient, given by  Eq. (\ref{B3}) of  \ref{AppB}, we obtain for $n_k,m_k, S_{kl}\ge1$:
\begin{eqnarray}
&&\mathcal{ P}(\{S_{kl}\}|\{n_k,m_l\})= (2\pi N)^{-\frac{(M-1)^2}{2}}
\left[\frac{ \prod_{k=1}^M \frac{n_k}{N} \frac{m_k}{N} }{ \prod_{k,l=1}^M p_{kl}}\right]^{\frac12}
\nonumber\\
&&\qquad\times\exp\left\{-N\mathcal{I}\left(\{p_{kl}\}\right)\right\}(1+\mathcal{O}(N^{-1}),
\label{E6}
\end{eqnarray}
where we have denoted by $\mathcal{I}$ the mutual information function, namely
\begin{eqnarray}
\mathcal{I}\left(\{{S_{kl}}/{N}\}\right) &\equiv& \mathcal{H}(\{n_k/N\}) + \mathcal{H}(\{m_k/N\}) - \mathcal{H}(\{S_{kl}/N\})
\nonumber\\
&=&\sum_{k,l=1}^M p_{kl}\ln\left(\frac{p_{kl}}{(n_k/N)(m_l/N)}\right),
\label{E7}\end{eqnarray}
with the Shannon entropy function denoted by  $\mathcal{H}$.

The sum in Eq. (\ref{E3}) can be replaced by an integral as $N\to \infty$, since the difference between the elements of two neighboring $p$-matrices  (the $N^{-1}$-scaled contingency tables $S_{kl}$)   is of order $1/N$. The Kronecker  delta symbols must be  replaced by the $N^{-1}$-scaled  Dirac delta functions:
\begin{eqnarray}
&&\Delta p_{kl}\sim \frac{1}{N} \to dp_{kl},\nonumber\\
&& \delta_{\sum_{l=1}^M S_{kl},n_k} \to \frac{1}{N}\delta\left(\sum_{l=1}^M p_{kl}-\frac{n_k}{N}\right),
\nonumber\\
&& \delta_{\sum_{k=1}^M S_{kl},m_l} \to \frac{1}{N}\delta\left(\sum_{k=1}^M p_{kl}-\frac{m_l}{N}\right).
\label{E8}\end{eqnarray}
There are only $2M-1$ independent constrains in the product  of $2M$ Kronecker deltas in Eq. (\ref{E3}),  since the  contingency table elements $S_{kl}$ sum to $N$,  giving both   the sum of the row sums and of the column sums. Hence,  the integration domain is $(M-1)^2$-dimensional. Using these observations and some elementary algebra we obtain from Eqs. (\ref{E3}), (\ref{E6}), and (\ref{E8}):
\begin{eqnarray}
&&\per(U[n_1,\ldots,n_M|m_1,\ldots,m_M]) \approx N!\left(\frac{N}{2\pi} \right)^{\frac{(M-1)^2}{2}}{\sqrt{ \prod_{k=1}^M \frac{n_k}{N} \frac{m_k}{N} }}
\nonumber\\
&&\times\int d\mu(\{p_{kl}\})
 \frac{\exp\left\{-N\left[\mathcal{I}(\{p_{kl}\})-\sum_{k,l=1}^Mp_{kl}\ln U_{kl}\right]\right\}}{\sqrt{\prod_{k,l=1}^Mp_{kl}}}.\qquad
\label{E9} \end{eqnarray}
Here we have introduced an integration measure $d\mu(\{p_{kl}\})$ over  an $(M-1)^2$-dimensional subspace in the convex set of all matrices with positive elements constrained only  by  the $2M-1$ Dirac delta functions from Eq. (\ref{E8}).\footnote{Note that an arbitrary subset of $2M-1$ delta functions can be used, see also   \ref{AppC}.} It reads

\begin{eqnarray}
d\mu(\{p_{kl}\}) &=& \left[\prod_{k,l=1}^Mdp_{kl}\right]\left[\prod_{k=1}^M\delta\left(\sum_{l=1}^M p_{kl}-\frac{n_k}{N}\right)\right]
\nonumber\\
&\times&\left[\prod_{l=1}^{M-1}\delta\left(\sum_{k=1}^M p_{kl}-\frac{m_l}{N}\right)\right].
\label{E10}\end{eqnarray}

The error of the approximation in   Eq. (\ref{E9}) is estimated  to have a multiplicative order  $\sim 1/N$, since this is the order of our approximation of the Fisher-Yates distribution by a smooth function in  Eq. (\ref{E6}), whereas replacing a finite sum by an integral brings also an error on the order of  difference between the values of two nearest lattice points, i.e. $\Delta p_{kl} \sim 1/N$.  The multidimensional  integral in Eq. (\ref{E9}) is in the  standard  form used for  asymptotic expansion in powers of $1/N$ by the saddle point  method (called also the steepest descent method). However, since  the integral representation  in Eq. (\ref{E9}) has already an error  of order  $ 1/N$, only the leading term of  the resulting asymptotic expansion is  meaningful.

\subsection{The matrix scaling problem giving the saddle points}

At this stage, let us recall the general formula for the  leading term of an $n$-dimensional integral,  given by the saddle point approximation, when the saddle points are simple \cite{FB}:
\be
\int d^n z e^{-N\phi(z)}g(z) \approx \left(\frac{2\pi}{N}\right)^\frac{n}{2}
\sum_{j} \frac{ \exp\{-N\phi(z_j)\} }{ \sqrt{ \mathrm{det} \left( \frac{ \partial^2\phi(z_j) }{\partial z^2 }\right) } }g(z_j),
\en{E14}
where the summation is over contributing saddle points $z_j$. The determinant in the denominator of Eq. (\ref{E14}) is of the Hessian matrix, i.e. the matrix composed of  the second-order derivatives of $\phi(z)$, taken at the respective  saddle point. The saddles $z_j$ are found  by a  deformation, as allowed by  analyticity of $\phi(z)$, of the integration domain in the extended complex-valued  space of $z$, such that the deformed domain is  contained in  the steepest descent  regions of the integrand.  The next term in the asymptotic expansion, as compared to the leading term of Eq. (\ref{E14}), has the relative  order of $1/N$.
Let us now apply the result (\ref{E14}) to our specific case and derive the saddle point approximation of the permanent. 

\subsubsection{The  matrix scaling problem}

The saddle points (matrices $p$, in our case) are found as extremals of the multivariate function  in the exponent of the integrand. In our case the function reads
\be
\phi = \mathcal{I}(\{p_{kl}\}) - \sum_{k,l=1}^Mp_{kl}\ln U_{kl}
\en{phi}
 with, however,   only $(M-1)^2$ independent variables out of the total $M^2$ matrix elements $p_{kl}$.  Using the Lagrange multipliers $\lambda_{k}$ and $\mu_l$  one can equivalently look for extremals of  the augmented function 
\be
\mathcal{F} \equiv \mathcal{I}(\{p_{kl}\}) - \sum_{k,l=1}^Mp_{kl}\ln U_{kl} - \sum_{k,l=1}^M\left(\lambda_k+\mu_l\right)p_{kl} .
\en{E11}
Equating the differential of $\mathcal{F}$ to zero we obtain that    the saddle points have the following general form
\be
p_{kl} = x_kU_{kl}y_l,
\en{E12}
where  the complex parameters $x_k$ and $y_l$ are determined by satisfying the margins imposed  on  $p$ by the Dirac delta functions in Eq. (\ref{E10}). In other words,  the diagonal matrices $X \equiv \mathrm{diag}(x_1, \ldots, x_M)$ and $Y\equiv \mathrm{diag}(y_1, \ldots, y_M)$  solve the  matrix scaling problem for  unitary matrix $U$: the scaled matrix $XUY$ must have  row and column sums specified by  boson distributions in the input and output modes, i.e.
\be
\sum_{l=1}^M x_kU_{kl}y_l = \frac{n_k}{N},\quad \sum_{k=1}^M x_kU_{kl}y_l = \frac{m_l}{N}.
\en{E13}

\subsubsection{Calculation of the  Hessian. The main result}

In our case,  the Hessian matrix is   with respect to some independent  $(M-1)^2$ variables  from the $M^2$ elements of  matrix $p$. Therefore,  an extension  of the saddle point method to the constrained integration is needed, which runs as follows. We rewrite the constrains on  variables $p_{kl}$ in Eq. (\ref{E10}) as a set of linear equations  by introducing a matrix $C_{j,kl}$, where the enumeration order of the double index $(k,l)$ is as follows $(k,l) = \{(1,1),\ldots,(1,M),(2,1),\ldots (2,M),\ldots,(M,1),\ldots,(M,M)\}$, i.e.  index $k$ runs slower  than    index $l$. The constraints can be  rewritten as follows\footnote{The specific subset of $2M-1$ constraints is  in accord  with  the selected measure in Eq. (\ref{E10}).}
\begin{eqnarray}
&& \sum_{(k,l)}C_{j,kl}p_{kl}= c_j, \qquad c_j \equiv \left( \frac{n_1}{N},\ldots,\frac{n_M}{N},\frac{m_1}{N},\ldots,\frac{m_{M-1}}{N} \right),
\nonumber\\
&& C_{j,kl} \equiv \left\{  \begin{array}{cc} \delta_{j,k}, & 1\le j\le M, \\
\delta_{j-M,l}, & M+1\le  j \le 2M-1.\end{array}\right.
\label{E15}\end{eqnarray}
Matrix $C$ in Eq. (\ref{E15}) has rank equal to $2M-1$.  It can be partitioned  into a $(2M-1)\times(2M-1)$-dimensional   nonsingular  submatrix  $C^{(I)}$  and a submatrix  $C^{(II)}$. These two matrices induce a similar  partition  of  the elements $p_{kl}$, treated  as a vector   with double index $(k,l)$. Using a vector notation $\bar{p}$, we can  cast  the system of constraints given by  Eq. (\ref{E15}) as follows
\be
C^{(I)}\bar{p}^{(I)} + C^{(II)}\bar{p}^{(II)} = \bar{c}.
\en{E16}
Eq. (\ref{E16}) allows to   extract    independent variables from the $M^2$ elements of $p$ and calculate the needed Hessian. 
First, by introducing two vectors, $\bar{\xi}$ consisting  of $2M-1$ dependent  integration variables and $\bar{\eta}$  of $(M-1)^2$ independent ones, as follows $\bar{\xi} = C^{(I)}\bar{p}^{(I)}$ and $\bar{\eta} = \bar{p}^{(II)}$, we satisfy  the  constraints by fixing  the value of $\bar{\xi}$ according to Eq. (\ref{E16}) (i.e. by integrating over $\bar{\xi}$ using the Dirac delta functions in Eq. (\ref{E10})) and obtain the rest of the measure  $d\mu$ as follows
\be
d\mu^\prime(\{p_{kl}\}) = |\mathrm{det}C^{(I)}|^{-1}\left[\prod_{j=1}^{(M-1)^2}d\eta_j\right].
\en{E17}
Second, due to linearity of  constraints (\ref{E16}), the determinant of the matrix of second derivatives of $\phi(\{p_{kl}\})$ with respect to $(M-1)^2$ independent variables can be   evaluated from the full matrix of second derivatives with respect to all variables $p_{kl}$ by using Eq. (\ref{E16}).    We get the following result
\begin{eqnarray}
&&\mathrm{det}\left(\frac{\partial^2 \phi}{\partial ( \bar{p}^{(II)} )^2 }\right) = \mathrm{det}\left\{[\widetilde{B},I]\left(\frac{\partial^2 \phi}{\partial \bar{p}^2 }\right)\left[\begin{array}{c} B\\ I \end{array}\right]\right\},
\nonumber\\
&& B \equiv -\left(C^{(I)}\right)^{-1}C^{(II)}.
\label{E18}\end{eqnarray}
Here  $[\widetilde{B},I]$ stands for  the block matrix constructed from  the transposed $(2M-1)\times(M-1)^2$-dimensional matrix $B$ and the $(M-1)^2\times(M-1)^2$-dimensional matrix unit $I$. Furthermore, the determinant on the r.h.s. of Eq. (\ref{E18}) can be further simplified by using
an  identity which generalizes  Sylvester's identity for  determinant of a block matrix (see \ref{AppC} for details) valid for a nonsingular matrix $A$:
\be
\mathrm{det}\left(C^{(I)}\right)^2 \mathrm{det}\left\{[\widetilde{B},I]A \left[\begin{array}{c} B\\ I \end{array}\right]\right\}
=\mathrm{det}(A) \mathrm{det}\left(CA^{-1}\widetilde{C}\right),
\en{E19}
where matrix $B$ is as in Eq. (\ref{E18}). In  our case   the Hessian matrix $A$ in the second differential of $\phi = \mathcal{I}(\{p_{kl}\}) - \sum_{k,l=1}^Mp_{kl}\ln U_{kl}$  is diagonal, i.e.
\be
d^2 \phi(\{p_{kl}\})  = \sum_{k,l=1}^M\frac{1}{p_{kl}} \left(dp_{kl}\right)^2,
\en{E20}
thus Eq. (\ref{E19})  applies  for $p_{kl}\ne0$. Note that the exponent with a large parameter $N$  in the integral on the r.h.s. of Eq.~(\ref{E9})  evaluated at a saddle point $p$ such that some $p_{kl}=0$ would be  infinite (thus this case  is ruled out). Indeed,  using   Eq. (\ref{E13}), we get at a  saddle point $p_{kl} = x_kU_{kl}y_l$:
\be
\exp\left\{ -N\left(\mathcal{I}(\{p^{(s)}_{kl}\})-\sum_{k,l=1}^Mp^{(s)}_{kl}\ln U_{kl}\right)\right\}
=  \prod_{k=1}^M\left(\frac{n_k}{Nx_k}\right)^{n_k}\left(\frac{m_k}{Ny_k}\right)^{m_k}.
\en{E21}

Now, using Eqs. (\ref{E14}),  and (\ref{E17})-(\ref{E21})  into Eq. (\ref{E9}) and noticing that
\[
\mathrm{det}\left(\frac{\partial^2 \phi}{\partial \bar{p}^2 }\right) =  \prod_{k,l=1}^M\frac{1}{p_{kl}},
\]
we obtain  a formula for the leading term approximation to the matrix permanent in the case of simple saddle points (our main result)
\be
\per(U[n_1,\ldots,n_M|m_1,\ldots,m_M]) \approx N!  \sqrt{ \prod_{k=1}^M \frac{n_k}{N} \frac{m_k}{N} }\sum_{s}\frac{\prod_{k=1}^M\left(\frac{n_k}{Nx^{(s)}_k}\right)^{n_k}\left(\frac{m_k}{Ny^{(s)}_k}\right)^{m_k}}{\sqrt{\mathrm{det}(D^\prime(p^{(s)}))}}.
\en{E22}
Here the sum over all contributing saddle points $p^{(s)}_{kl}=x^{(s)}_k U_{kl}y^{(s)}_l$ is implied and matrix $D^\prime$ in the denominator is as follows
\be
D^\prime \equiv C \left(\frac{\partial^2 \phi}{\partial \bar{p}^2 }\right)^{-1}\widetilde{C},
\en{E23}
with the matrix $C$  given by Eq. (\ref{E15}).  The sparsity of  $C$  allows one  to easily find the   explicit form of the $(2M-1)\times(2M-1)$-dimensional matrix $D^\prime$.  We get that   $\mathrm{det}(D^\prime)$, appearing in the denominator on the r.h.s. of Eq. (\ref{E22}), is actually equal to any of  $(2M-1)\times(2M-1)$-dimensional  principal minors (i.e. obtained by crossing out the same column and row) of the full  $2M\times2M$-dimensional  matrix $D$, defined as follows
\be
D  = \left(\begin{array}{ccc|ccc}
\frac{n_1}{N}  &  &  0 & & & \\
& \ddots & & &p&  \\
0 &  & \frac{n_M}{N} & & &\\
& & & & & \\ \hline
& & &\frac{m_1}{N} & & 0 \\
& \widetilde{p} &  &  &\ddots & \\
 & & &0 & & \frac{m_M}{N}
\end{array}\right)
\en{E24}
(in  Eq. (\ref{E24})  we have used guiding lines to emphasize the block structure of $D$). Any of the above specified  principal minors can be used due to the  fact that any subset of $2M-1$ constraints from the full set of $2M$  ones could be used in the definition of the integration measure $d\mu$ in Eq.~(\ref{E10}).   The determinant $\mathrm{det}(D^\prime)$   can be   reduced to a simpler form (see   \ref{AppC}, where  the  above described property of the principal minors of $D$ is also directly verified). We have
\begin{eqnarray}
\mathrm{det}(D^\prime)& =&\left[ \prod_{k=1}^M\frac{n_k}{N}\right]\mathrm{det}\left( \Lambda^\prime_2 - \widetilde{p}^\prime \Lambda^{-1}_1p^\prime\right)
\nonumber\\
&=&\left[ \prod_{k=1}^M\frac{m_k}{N}\right]\mathrm{det}\left( \Lambda^\prime_1 - p^\prime\Lambda^{-1}_2\widetilde{p}^\prime \right),
\label{DET}\end{eqnarray}
where we have denoted $\Lambda_1 = \mathrm{diag}(\frac{n_1}{N},\ldots,\frac{n_M}{N})$ and $\Lambda_2 = \mathrm{diag}(\frac{m_1}{N},\ldots,\frac{m_M}{N})$ with  matrices $\Lambda_{1,2}^\prime$ and $p^\prime$  taken from $D^\prime$ (i.e. the   matrix $D$  of Eq. (\ref{E24}) with one row and column with the same index crossed  out).

The  above discussion implies that the symmetries of   bosonic probability amplitudes are preserved\footnote{This is also manifested  by \textit{exact} cancellation of  probability amplitudes in the  generalized HOM effect, Figs.~3 and 6 of section \ref{sec3}.}
by the saddle point  approximation.  For instance, the inversion symmetry: ${}_g\langle m_{1},\ldots,m_{M}|n_{1},\ldots,n_{M}\rangle_f  = {}_f\langle n_{1},\ldots,n_{M}|m_{1},\ldots,m_{M}\rangle_g^*$.

\subsection{Comparison with classical identical particles on a network }

Let us compare the transition probabilities of  indistinguishable bosons   with the transition probabilities of classical particles (which we   consider identical). In the classical case, the elementary process of Fig.~2 means  redistribution of $n_k$ identical classical particles from the $k$th input mode into the  $M$ output modes with the output distribution given by the same contingency table $S_{kl}$. The difference is that the probabilities are multiplied and summed up, thus instead of the amplitude of an elementary quantum  process, given by the product $\prod_{l=1}^MU_{kl}^{S_{kl}}$  in Fig.~2, we have  the probability of an elementary classical process, given by $\prod_{l=1}^M |U_{kl}|^{2S_{kl}}$.  As the particles are identical (i.e. the paths of the individual particles through the network are not traced), the  total probability of such an elementary process is given by the latter product multiplied by the number of redistributions of $n_k$ identical particles from the $k$th input mode into $M$ output modes, i.e. by the factor $\frac{n_k!}{\prod_{l=1}^MS_{kl}!}$. Summing up over all such  probabilities (i.e. over  the elementary  processes from different  input modes) and  identifying   the Fisher-Yates distribution in the summation, we obtain the transition probability of $N$ classical particles through $M$-mode network   (see also Ref.~\cite{MPI}) 
\begin{eqnarray}
P(n_1,...,n_M|m_1,...,m_M)& =&\frac{N!}{\prod_{k=1}^Mm_k!}\left\langle \prod_{k,l=1}^M|U_{kl}|^{2S_{kl}}\right\rangle\nonumber\\
&=&\frac{\mathrm{per}(|U|^2[n_1,\ldots,n_M|m_1,\ldots, m_M])}{\prod_{k=1}^Mm_k!},
\label{Class}\end{eqnarray}
where  the input and output distributions are  $\{n_1,\ldots,n_M\}$ and $\{m_1,\ldots,m_M\}$, respectively    (cf. with the quantum probability amplitude  given by  Eqs. (\ref{E1}) and (\ref{E5})). For instance, for Bell multiports $|U_{kl}|^2 = \frac{1}{{M}}$ and we obtain the resulting probability as follows (see also Ref.~\cite{ZTL})
\be
P(n_1,...,n_M|m_1,...,m_M) = \frac{N!}{M^N\prod_{k=1}^Mm_k!}. 
\en{ClassBell}

One can apply the saddle point approximation also to the classical probability given by Eq. (\ref{Class}).  Using this simple observation, we can compare complexity of the saddle point approximation   in the  classical and  quantum cases.  A very important difference is spotted immediately: the classical analog of the matrix scaling problem  is formulated for a matrix of positive elements $A_{kl} \equiv |U_{kl}|^2$ (note that  matrix $A$ is doubly stochastic, i.e. its row and column sums are equal: $\sum_{k=1}^MA_{kl} = \sum_{l=1}^MA_{kl} = 1$).  It is known that the matrix scaling problem for a positive matrix has a unique \textit{positive} solution, since it is equivalent to a minimization problem of a convex function \cite{MatScal,MatScal2,MScalUnique}.  Note that  the corresponding  saddle point belongs to the integration  domain over the contingency tables $p_{kl} = S_{kl}/N$, i.e. $x_kA_{kl}y_l<1$, since both $A_{kl}\ge0$ and  $x_k,y_k>0$ (the positive solution is constrained by the margins). Therefore,   it is the only contributing saddle point in the classical case.  This is  remarkably different from the quantum case, where, as  is discussed below, generally there are more than one \textit{complex-valued} contributing saddle points. They loose interpretation of the dominating ``real processes" of Fig.~2 (since the corresponding contingency table $S_{kl}$  is complex). However, the saddle points  describe in a simpler way the    quantum interferences between exponentially many of  such real processes in Eq. (\ref{E5}).

Finally,  the saddle point method   \textit{ reproduces the exact result} for  the classical analog of  Bell multiports, i.e.   for classical particles on an equal probabilities network $|U_{kl}|^2 = \frac{1}{M}$. Indeed, let us see that the saddle point approximation (\ref{E22}) with $U_{kl}$ replaced by $|U_{kl}|^2$  reproduces Eq. (\ref{ClassBell}) if substituted into Eq. (\ref{Class}). In this special case of a network matrix, the solution to the matrix scaling problem   (\ref{E13})  can be found explicitly for any network size $M$: $x_k = \sqrt{M}\frac{n_k}{N}$ and $y_k = \sqrt{M}\frac{m_k}{N}$. Thus the only contributing saddle point reads $p_{kl} = \frac{n_k}{N}\frac{m_l}{N}$, which we rewrite as $p = |\frac{n}{N}\rangle\langle \frac{m}{N}|$ (i.e. adopting the vector-column $|\cdot\rangle$ and vector-row $\langle\cdot |$ notations). Calculation of the determinant in Eq. (\ref{DET}) reduces in this case to applying  Sylverster's determinant identity ($ \mathrm{det}(I_m - A_{m,n}B_{n,m}) = \mathrm{det}(I_n - B_{n,m}A_{m,n})$):
\begin{eqnarray}
\mathrm{det}(D^\prime)& =&\left[ \prod_{k=1}^M\frac{n_k}{N}\right]\mathrm{det}\left( \Lambda^\prime_2 - \widetilde{p}^\prime \Lambda^{-1}_1p^\prime\right)
\nonumber\\
& = & \left[ \prod_{k=1}^M\frac{n_k}{N}\right]\left[ \prod_{k=1}^{M-1}\frac{m_k}{N}\right]\mathrm{det}\left(I_{M-1} -(\Lambda_2^\prime)^{-1}\left|\frac{m}{N}^\prime\right\rangle\left\langle\frac{m}{N}^\prime\right|\right)\nonumber\\
& = & \left[ \prod_{k=1}^M\frac{n_k}{N}\right]\left[ \prod_{k=1}^{M-1}\frac{m_k}{N}\right]\left(1 -\left\langle\frac{m}{N}^\prime\right|(\Lambda_2^\prime)^{-1}\left|\frac{m}{N}^\prime\right\rangle\right) = \left[ \prod_{k=1}^M\frac{n_k}{N}\frac{m_k}{N}\right]\!. \quad
\label{Cldet}\end{eqnarray}
By using the explicit form of the saddle point,  substituting Eq. (\ref{Cldet}) into Eq. (\ref{E22}) and the resulting expression   into Eq.~(\ref{Class}) one recovers the exact result given by Eq.~(\ref{ClassBell}) for a classical analog of  Bell multiports.  

\section{Testing  accuracy  of the saddle point approximation}
\label{sec3}

To apply the saddle point approximation (\ref{E22}) we first  have to solve the matrix scaling problem (\ref{E13}) which is  a bilinear system of equations in $x$ and $y$.   One can reduce the number of variables by half by resolving one of the equations  in Eq. (\ref{E13}), for instance, $y_l = \sum_{k=1}^M U^*_{kl}({n_k}/{Nx_k})$.  By introducing a  vector $R$ containing all  $M-1$ independent variables\footnote{Multiplication of  all $x$-variables by a  complex number $\lambda$, $x_k\to \lambda x_k$, does not change the saddle points, since it induces the inverse  scaling of the $y$-variables: $y_k\to y_k/\lambda$.} and  a set of  $M-1$  vectors $Z^{(l)}$, defined  as follows:
\be
R_k \equiv \sqrt{\frac{n_M}{n_k}}\frac{x_k}{x_M}, \qquad Z^{(l)}_k  \equiv \sqrt{\frac{n_k}{N}}U_{kl}, \quad k = 1,\dots,M,
\en{matrA}
we obtain from Eq.~(\ref{E22}) a reduced   system    in  the following form 
\be
\left(\sum_{k=1}^M R_k Z^{(l)}_k\right)\left(\sum_{q=1}^M\left[Z^{(l)}_q\right]^*R^{-1}_q \right)= \frac{m_l}{N}, \qquad l = 1,\ldots,M-1.
\en{E26}
Eq. (\ref{E26})  is  hard  to solve analytically  for more than two modes (despite  considerable efforts, the  solution has not been found even for the simple case of Bell multiports).  Moreover, \textit{all}  solutions of Eq. (\ref{E26}) are needed, since all saddle points  contribute to the approximation in general. On the other hand, it is relatively easy to find solutions to   Eq.~(\ref{E26}) numerically (for instance, by   the all-purpose nonlinear equations solver available in MATLAB  with a random  initial guess to find all possible solutions). It was found that the total number of solutions is dependent on    $\{n_1/N,...,n_M/N\}$ and $\{m_1/M,...,m_M/N\}$ and that there can be symmetries leading to degeneracies.   The absence of a  formula for   the total number of solutions  prevents analysis of the computational complexity   of Eq. (\ref{E26}).

Below we consider  accuracy of the saddle point approximation in two cases: the beam splitter and the tritter, where the beam splitter allows an analytical solution, while already the tritter case requires a numerical solution.

\subsection{The two-mode (beam splitter) case}
\label{sec3.1}

 The case of beam-splitter, $M=2$,  is  analytically solvable.  As is known \cite{Berns},  two and three  dimensional networks are uniquely defined by  moduli $|U_{kl}|$ of the unitary matrix elements, whereas the $2M-1$  phases are scaled out by changing the unimportant phases of the  input and output  states. Let us  consider the  symmetric beam splitter, which is given  by the following matrix 
\be
U = \left(\begin{array}{cc} -\frac{1}{\sqrt{2}} &  \frac{1}{\sqrt{2}}  \\ \frac{1}{\sqrt{2}}  & \frac{1}{\sqrt{2}} \end{array} \right).
\en{E27}
Then Eq. (\ref{E26}) leads to a  quadratic equation for $R_1 = \frac{\sqrt{n_2}}{\sqrt{n_{1}} }\frac{x_{1}}{x_2}$:
\be
R_1^2 - 2\gamma{R_1}+1 =0, \quad \gamma = \frac{ m_2-m_1}{2\sqrt{n_1n_2}}.
\en{E28}

\subsubsection{The saddle points} 

There are two saddle points $p$ whose $x$  component in Eq. (\ref{E12}) reads\footnote{To obtain  $x_{1,2}$ from $R_1$ we have taken  into account the scale invariance  $x\to\lambda x$ and $y\to y/\lambda $.}
\begin{eqnarray}
&& x_1 = \sqrt{\frac{n_1}{N}}e^{i\phi/2}, \quad x_2 = \sqrt{\frac{n_2}{N}}e^{-i\phi/2},\nonumber\\
&& e^{i\phi} \equiv  \gamma \mp i\sqrt{1-\gamma^2}.
\label{E29}
\end{eqnarray}
The $y$ component is  given by $y_1 = ( -\frac{n_1}{Nx_1} + \frac{n_2}{Nx_2})/\sqrt{2}$ and  $y_2 = ( \frac{n_1}{Nx_1} + \frac{n_2}{Nx_2})/\sqrt{2}$. In Eq. (\ref{E29}) we have introduced a phase $\phi$, however, it is a real value only under the condition that $\gamma^2<1$, i.e. when
\be
(\Delta n)^2 +  ( \Delta m )^2  < N^2,
\en{E30}
with $\Delta n = n_2-n_1$ and $\Delta m = m_2 - m_1$. Under  condition (\ref{E30}) the $y$ component can  be given as follows
\begin{eqnarray}
&& y_1 = \sqrt{\frac{m_1}{N}}e^{i(\delta+\psi)/2}, \quad y_2 = \sqrt{\frac{m_2}{N}}e^{i(\delta-\psi)/2}, \nonumber\\
&& e^{2i\psi} \equiv \sigma \mp i \sqrt{1- \sigma^2}, \quad \sigma \equiv \frac{\Delta n}{2\sqrt{m_1m_2}},\nonumber\\
&& e^{i\delta} \equiv \frac{i\Delta n\Delta m \pm \sqrt{N^2 - (\Delta n)^2 -(\Delta m)^2}}{\sqrt{n_1n_2m_1m_2}}.
\label{E31}\end{eqnarray}
Since $4m_1m_2(1-\sigma^2 ) = 4n_1n_2(1- \gamma^2) = N^2 - (\Delta n)^2 - (\Delta m)^2$,  the same threshold condition (\ref{E30}) applies to phases $\psi$ and $\delta$.  When condition (\ref{E30}) is violated, the phases $\phi$, $\psi$, and $\delta$  become complex-valued ($\phi$ and $\psi$ become imaginary). This corresponds to a phase transition in  the  bosonic probability amplitudes  (see below).

The  saddle points   are given   by Eq. (\ref{E12}). We get:
\begin{eqnarray}
&& p_{11} = \frac12\left( \frac{n_1}{N} +\sqrt{\frac{n_1n_2}{NN}}\left[-\gamma \pm i\sqrt{1-\gamma^2}\right]\right),\label{p11}\\
&&   p_{12} = \frac12\left( \frac{n_1}{N} +\sqrt{\frac{n_1n_2}{NN}}\left[\gamma \mp i\sqrt{1-\gamma^2}\right]\right),\label{p12}\\
&& p_{21} = \frac12\left( \frac{n_2}{N} +\sqrt{\frac{n_1n_2}{NN}}\left[-\gamma \mp i\sqrt{1-\gamma^2}\right]\right),\label{p21}\\
&&   p_{22} = \frac12\left( \frac{n_2}{N} +\sqrt{\frac{n_1n_2}{NN}}\left[\gamma \pm i\sqrt{1-\gamma^2}\right]\right),\label{p22}
\label{p}\end{eqnarray}
valid in  the  whole  domain $|\Delta n|\le N$, $|\Delta m|\le N$. When  condition (\ref{E30}) is satisfied, i.e. when $|\gamma|\le1$, both saddle points are complex-valued and contribute to the integral in Eq. (\ref{E22}). When it is  violated, the saddle points become real valued.  However, one of them  does not contribute to the integral, since  it escapes from  the integration domain $0\le p_{kl}\le1$. Which one of the saddle points contributes depends on the sign of $\gamma$, i.e. the sign of  $\Delta m$,  and the ratio $n_1/n_2$ (see also Fig.~3(c) below).

Finally, after a simple algebra, the determinant  given by  Eq. (\ref{DET}) becomes
\begin{eqnarray}
\mathrm{det}(D^\prime) =  &\mp&\frac{ 1}{8}e^{i\delta}\left(1 - \left[\frac{\Delta n}{N}\right]^2\right)^{\frac12}\left(1 - \left[\frac{\Delta m}{N}\right]^2\right)^{\frac12}
\nonumber\\
&\times& \left(1 - \left[\frac{\Delta n}{N}\right]^2-\left[\frac{\Delta m}{N}\right]^2\right)^{\frac12},
\label{E32}\end{eqnarray}
where we have used that $16n_1n_2m_1m_2=\left(N^2 - (\Delta n)^2\right)\left(N^2 - (\Delta m)^2\right)$.
Substituting Eqs. (\ref{E29}), (\ref{E31}), and (\ref{E32}) into Eq. (\ref{E22}) and the resulting  approximation into Eq. (\ref{E1}) we obtain the saddle-point approximation  to  bosonic probability amplitudes of the symmetric beam-splitter (\ref{E27}).

\subsubsection{Comparison with the exact result}

\begin{figure}
\begin{center}
\includegraphics[width=0.9\textwidth]{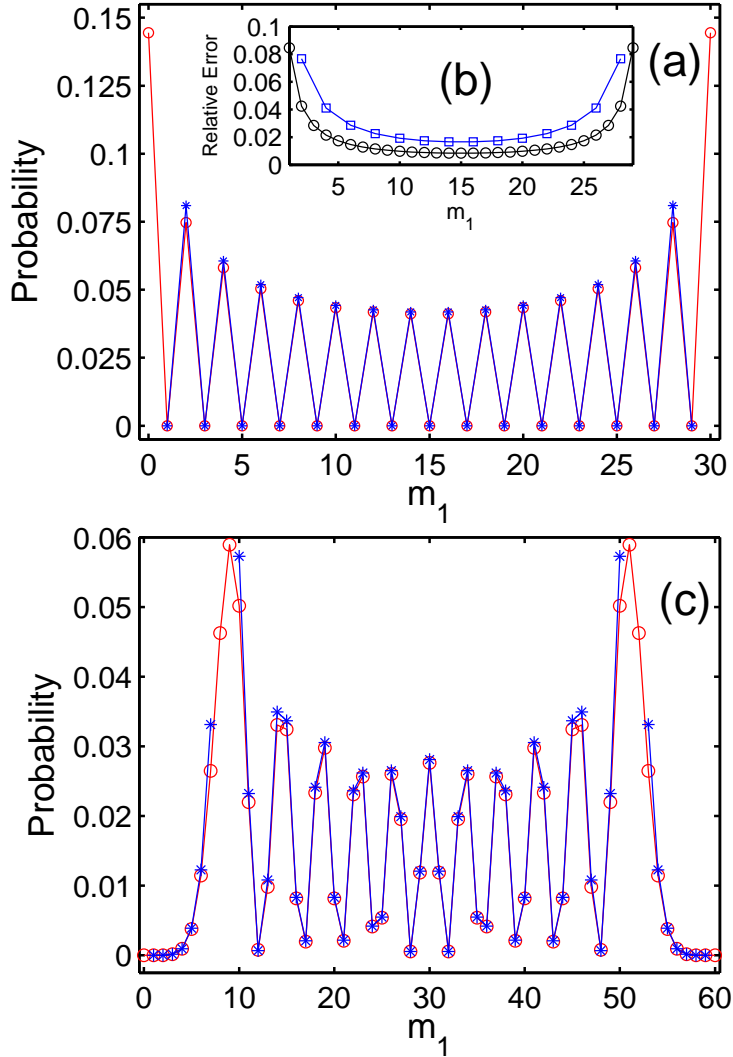}
 \caption{  Comparison of the saddle point approximation (shown by the stars) with the exact result (shown by the circles) for  bosonic probability amplitude of the beam splitter (\ref{E27}) (to guide the eye, the  numerical points are connected by lines). Panel (a): $N = 30$ and $n_1=15$. Panel (b) gives the relative error of panel (a) at the even points of $m_1$ (the squares) compared with the relative error of approximation of the binomial coefficient $( {N \atop  m_1})$ based on the Stirling  formula (the circles). Panel (c): $N=60$ and $n_1=10$, the two regions close to $m_1=10$ and $m_1=50$ are in the vicinity of the circle $(\Delta n)^2 +(\Delta m)^2 = N^2$ where the simple saddle point approximation of Eq.  (\ref{E22}) fails (diverges). For $10< m_1 <50$ the two saddle points contribute, while for $m_1<10$ or $m_1>50$ just one saddle point contributes. }
 \end{center}
\end{figure}

To compare with the exact result  the following expression  for bosonic probability amplitude for network matrix of Eq.  (\ref{E27})  will be used  (see also Ref.~\cite{LM})
\be
{}_{g}\langle m_1,m_2|n_1,n_2\rangle_{f} = \frac{\sqrt{n_1!n_2!m_1!m_2! }}{2^{N/2}}\sum_{q}\frac{(-1)^q}{q!(n_1-q)!(m_1-q)! (m_2+q-n_1)!}.
\en{E33}
The summation index $q$  satisfies $\mathrm{max}(0,n_1\!-m_2)\le q\le \mathrm{min}(n_1,m_1)$. 
\begin{figure}
\begin{center}
\includegraphics[width=0.85\textwidth]{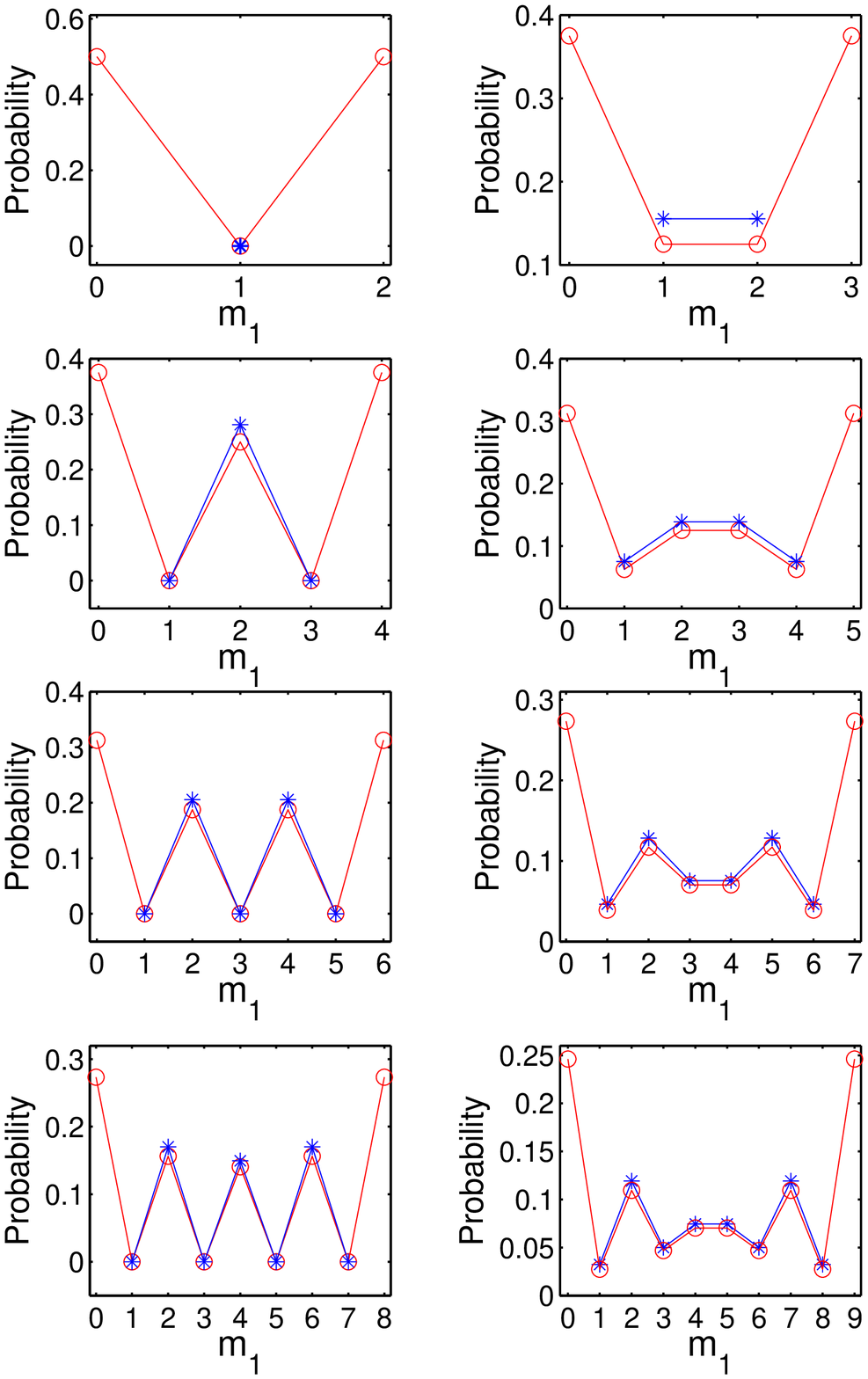}
 \caption{  The saddle point approximation for the beam splitter  (\ref{E27})  (shown by the stars)   is compared with the exact result (shown by the circles) for $N=\{2,4,6,8\}$, the left column from top to bottom, and for $N=\{3,5,7,9\}$, the right column from top to bottom. In the left column $n_1 = N/2$, while in the right one $n_1=(N+1)/2$ (note that the approximation is undefined for the endpoints $m_1=0$ and $m_1=N$). }
 \end{center}
\end{figure}

The correspondence of the exact result (\ref{E33}) with the saddle point approximation can  be divided into three regions. In the first region  condition (\ref{E30}) is  satisfied. This  region contains  the generalized HOM effect \cite{CST,LM}. In this case  the amplitudes of two contributing terms in Eq. (\ref{E22}), corresponding to  two saddle points,  have the same moduli and the only difference lies in  their relative phase. The corresponding domain in the two-dimensional plane  with coordinates $\Delta n$ and $\Delta m$ is the  inside of the circle  $(\Delta n)^2 +(\Delta m)^2 \lesssim N^2$.  There is cancellation of the probability amplitudes for $n_1=n_2 = N/2$ and odd values of $m_1$, Fig.~3(a) (see also Refs.~\cite{CST,LM}). In the saddle point approach, the cancellation is due to symmetry of the two saddle-point contributions with  the only difference being  their  relative  phase   given by $(-1)^{m_1}$ (note: the  cancellation is captured \textit{exactly} by the saddle-point approximation).

On the other hand, there is another regime: the exponential decay of the probability amplitude as $m_1$ approaches either $0$ or $N$, see Fig.~3(c). This regime has not been studied  previously (for instance, the approximation of Ref.~\cite{LM} only captures the oscillating regime). It  appears when  condition (\ref{E30}) is violated.  In this case, there is just one contributing saddle point, the one which has the smallest moduli contribution to the permanent (this is similar to what occurs in  the saddle-point approximation to the Airy function).

The third region is about the circle $ (\Delta n)^2 +  ( \Delta m )^2  \approx  N^2$. This region contains two  coalescing saddle points and cannot be  approximated by Eq.  (\ref{E22}) valid for the simple saddle points only (their contributions diverge on this circle, which is due to the determinant (\ref{E32}) approaching  zero).   Outside this region, which is restricted to  narrow neighborhoods of the points $m_1=10$ and $m_1=50$ in Fig.~3(c), the saddle point approximation has a very good accuracy as is shown in Fig.~3(b), where the accuracy of the approximation (\ref{E22}) is compared to that  for the  binomial coefficient, given by Eq. (\ref{B3}) of  \ref{AppB} and used to build the approximation of the Fisher-Yates distribution  (\ref{E6}). It is seen that the relative error is approximately twice as that in the approximation of the binomial coefficient. 

\begin{figure}
\begin{center}
\includegraphics[width=0.85\textwidth]{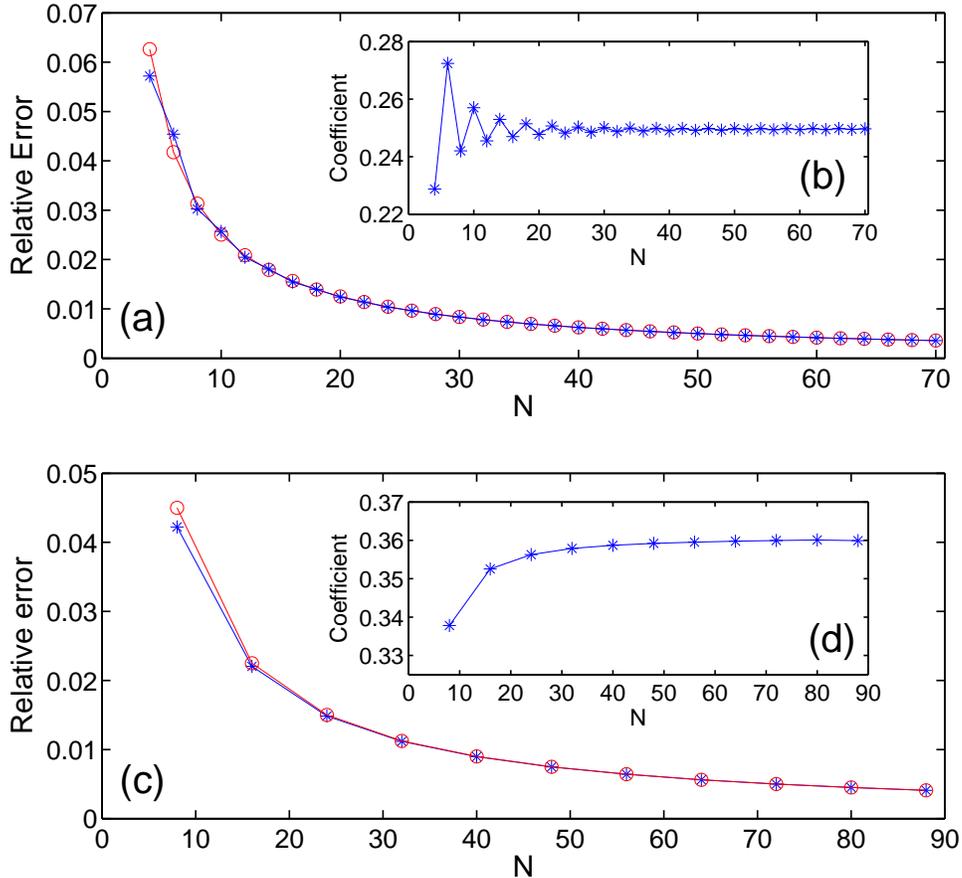}
 \caption{  The relative error $\mathcal{E} = \mathcal{E}(N)$  of the saddle point approximation  (\ref{E22}) (shown by the stars) for the beam splitter (\ref{E27})  for  $n_1=N/2$ and $m_1=N/2$, panels (a) and (b), and $n_1=3N/4$ and $m_1=N/2$, panels (c) and (d),  compared with the inverse proportionality law $f = \mathcal{E}(N_{max})N_{max}/N$ (shown by the circles), where $N_{max}$ is the  largest value of $N$.  To guide the eye the data are connected by lines. The insets (b) and (d)  give  the coefficient $C(N) = \mathcal{E}(N) N$.   }
 \end{center}
\end{figure}

\begin{figure}
\begin{center}
\includegraphics[width=0.85\textwidth]{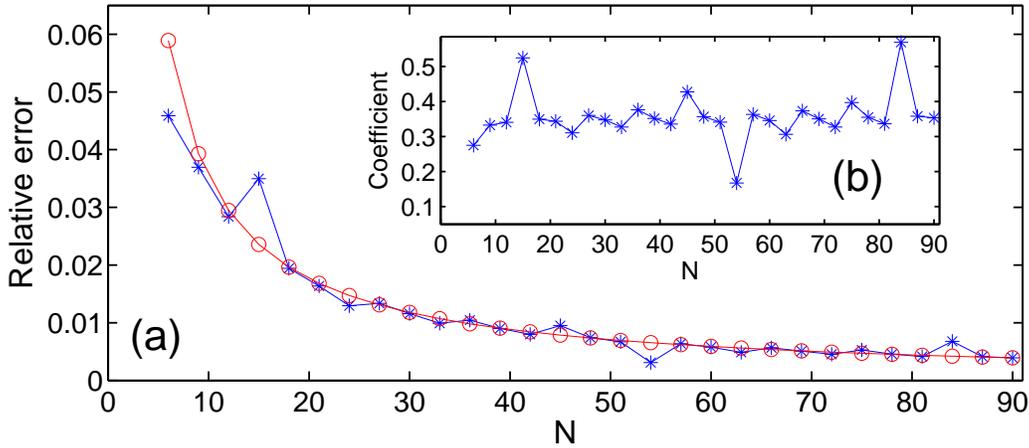}
 \caption{   Similar as in Fig.~5 but for  $n_1=2N/3$ and $m_1=N/3$ (the inverse proportionality law is fitted by using the data point at $N=90$).  }
 \end{center}
\end{figure}

The transition  from the two contributing saddle points, with an oscillating probability amplitude as function of $m_1$,  to a single contributing saddle point,  with an exponentially decaying probability amplitude,  is similar to the Airy function behavior, thus the related integral  in Eq. (\ref{E22})  can be, in principle, expressed  through a linear  combination of the Airy function and its first derivative. Such  results are available for the real-valued coalescing saddle points  (for instance, in Refs.~\cite{FB,FP}). However, the method  needs to be generalized to  the complex-valued case before it  could be applied to the integrals approximating  the bosonic probability amplitudes. 

Finally, it is interesting to observe that the saddle point approximation can give  correct results down to the very small number of bosons, for instance, it correctly predicts the  HOM effect \cite{HOM} (though, obviously,  small $N$ violate    the assumption  $N\gg1$). Several results for small number of bosons are collected in Fig.~4, where we have $2\le N\le 9$.

\subsubsection{Scaling of the relative error of the saddle point approximation}

Let us verify  that the relative error   scales as $1/N$ for fixed $n_1$ and $m_1$.   The scaling of the relative error of the saddle point approximation is given in Figs. 5 and 6.  Comparing   Fig.~5   with  Fig.~6 one can notice the  oscillations of the relative error around the law of  the inverse proportionality  in the latter case. The origin of these  oscillations is unclear. For instance, they are not due to approaching the boundary circle $ (\Delta n/N)^2  + (\Delta m/N)^2 =1$, since all data points from the same figure represent  one and the same point    in the $(\Delta n/N,\Delta m/N)$-square. The only explanation is a  very complicated  general dependence  of bosonic probability amplitude on $N$ for a fixed set of distributions $\{n_1/N,...,n_M/N\}$ and $\{m_1/M,...,m_M/N\}$ due to the fact that the  phases of the individual saddle point contributions are multiplied  by $N$.

\subsection{The three-mode (tritter) case}
\label{tritt} 

\begin{figure}
\begin{center}
\includegraphics[width=0.78\textwidth]{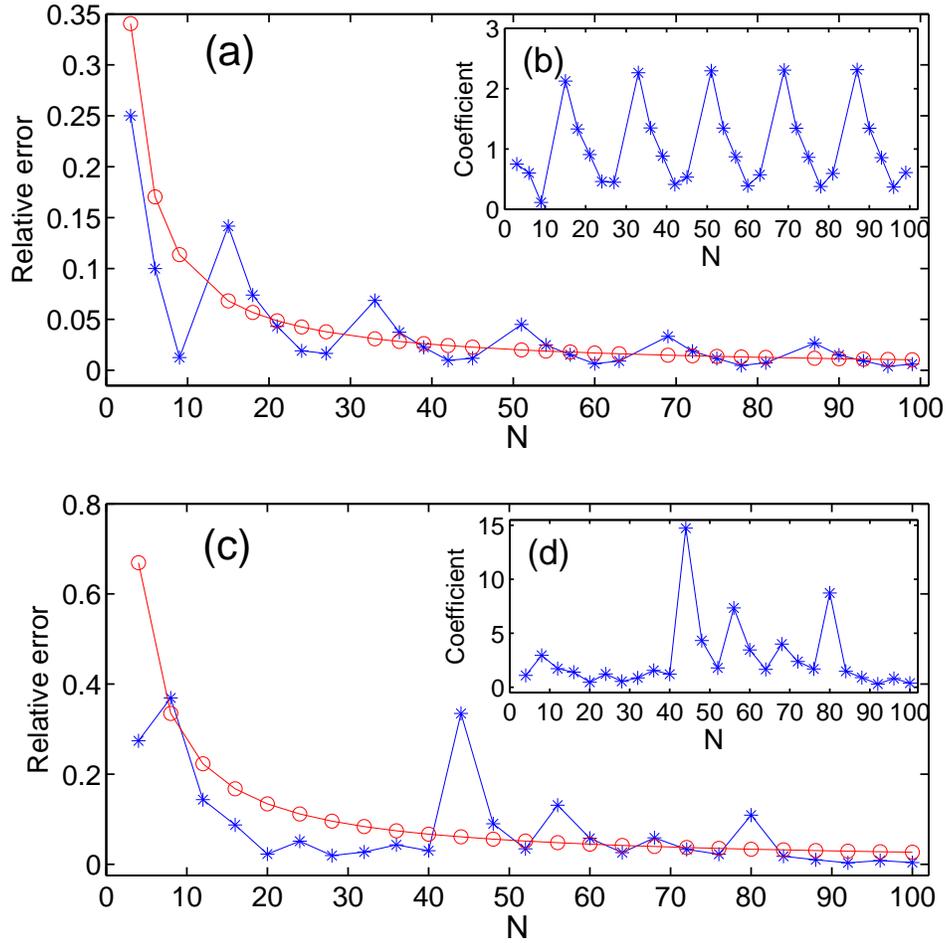}
 \caption{  The relative error $\mathcal{E} = \mathcal{E}(N)$  of the saddle point approximation  (\ref{E22})  for the  tritter (\ref{tritterU}) (shown by the stars) is compared with  the inverse proportionality law $f = \bar{C}/N$ (shown by the circles), where $\bar{C}$ is found by averaging $C(N) =\mathcal{E}(N)N$ over the numerical data points. To guide the eye the data are connected by lines.  Here $n_k/N = 1/3$ and $m_k/N = 1/3$ with $N(1) = 3$  in panel (a) and (b), while   $n_k/N = \{1/2, 1/4, 1/4\}$ and  $m_k/N = \{1/4, 1/2, 1/4\}$ with $N(1) = 4$ in panel (c) and (d). Panels (b) and (d) give the coefficient $C(N)$. }
 \end{center}
\end{figure}

Let us now consider three mode network (the tritter). The canonical symmetric tritter,  used, for instance,  in  the recent experiment \cite{HOM2zero}, has the following network matrix 
\be
U = \frac{1}{\sqrt{3}}\left( \begin{array}{ccc} 1 & 1 & 1 \\ 1 & e^{2i\pi/3} & e^{-2i\pi/3} \\ 1 & e^{-2i\pi/3} & e^{2i\pi/3}
\end{array}\right).
\en{tritterU}
Nonlinear system in Eq.~(\ref{E26}) with $U$ from Eq. (\ref{tritterU}) seems to be   unsolvable analytically, however, numerical solution contains at most six different saddle points. Moreover, numerical simulations with random three-mode unitary matrices $U$ has shown that six is the maximal number of saddle points  for any   three-mode network, where in most cases all saddle points contribute to the approximation and the respective vector parameters $x$ and $y$ have the following ``most probable" form  (after fixing one of the $x$-vector elements due to the scale invariance $x\to \lambda x$ of Eq. (\ref{E26}))
\be
x_k = \sqrt{\frac{n_k}{N}}e^{i\phi_k}, \qquad y_k = \sqrt{\frac{m_k}{N}}e^{i\psi_k}, \quad k = 1,2,3,
\en{saddles3} 
where $\phi_k$ and $\psi_k$ are real values (phases). This ``most probable" form  is very similar to  the general    solution  in the beam splitter  case of section \ref{sec3.1}. However, the corresponding contributions to the approximation from such saddle points are not always of the same moduli as distinct from  two-mode network. The approximation is not  even qualitatively correct for small number of bosons. Moreover, it is found that  the relative error always   has oscillations reminiscent of those in Fig.~6  (what can explain the poor performance of the approximation for  small $N$ in this case). Behavior of the relative error  is illustrated in Fig.~7. Computation of the  matrix permanent is carried out  by a modified Ryser's algorithm (similar as in Ref. \cite{Thesis}). Such  algorithm  has only polynomial in $N$ complexity   as is shown  in  Appendix D.


\section{Conclusion}
\label{sec4}

We have shown that an  asymptotic  evaluation of matrix permanents giving bosonic probability amplitudes in   unitary  linear  networks is possible for large number of bosons $N$ and  fixed network size $M$, such that  $N\gg M$.  The asymptotic approximation reduces   the problem of evaluation of permanents of  $N\times N$-dimensional  matrices with repeated rows and columns to a solution of a matrix scaling problem for $M\times M$-dimensional network matrix, which is a system of bilinear equations in  $2M-1$ variables.  For simple saddle points, an  explicit formula for bosonic probability amplitudes is derived.    
 
The asymptotic approximation  has been   compared with  the exact analytical result available   for  two-mode network, i.e. the beam-splitter, and has been found to have good accuracy correlated with   accuracy of  the approximation of   multinomial coefficient (used as a building block of the saddle-point approximation). The approximation error   is   studied also for   three-mode network, i.e. the tritter, where the saddle points were  found numerically.  Interestingly, in the  beam splitter case,  the approximation correctly reproduces  behavior of probability amplitudes even for small number of bosons, for instance, it reproduces the original HOM effect.  The relative error of the approximation is found to scale inversely with the number of bosons, however, the scaling is plagued by oscillations about the inverse scaling law  (the origin of which is unclear).   These oscillations degrade the approximation for small number of bosons in the case of  the tritter, where,  in contrast to the beam splitter case,  the approximation performs poorly for small number of bosons.

There are   various regimes of behavior of   bosonic probability amplitudes in unitary networks, which are dependent on the number of contributing saddle points.  For instance,   in the   beam-splitter case  there are two regimes: (i) the oscillating regime, when two saddle point contribute to bosonic probability amplitude  and  the generalized Hong-Ou-Mandel effects take place, and (ii) the regime of exponential decay of  bosonic  probability amplitudes, when only one saddle point contributes.  
 
Practical application of the method  is conditioned on  solution of the  matrix scaling problem giving the saddle points  (where the whole set of  solutions is generally required). Another problem of different type has to be solved before practical application of the method is attempted. One has to   derive a general formula for the saddle point method applicable when the saddle points coalesce.  There is also an important topological problem of identifying  the contributing saddle points, which is  a highly nontrivial one in general \cite{FB}, but may have general  solution for the type of integrals appearing in the saddle point approximation of matrix permanents.     
 
It is unlikely that there is analytical   solution to the matrix scaling   problem  for $M$-mode network with $M>2$, since the total number of different solutions grows rapidly with $M$. For $M=2$ there are at most two saddle points. Numerical simulations with random network matrices indicate that  for $M=3$  there are at most  six different saddle points, while for $M=4$ there are at most twenty different saddle points. Therefore, the matrix scaling problem may have an exponential in $M$ computational  complexity (note that this complexity is an attribute of a quantum network:  for classical particles on a network there is just one contributing saddle point).       
 
\section*{Acknowledgements}
The author  would like to thank the referee for many  valuable comments  that resulted in substantial  improvement of the presentation. This work was supported by the CNPq and FAPESP of Brazil.

\appendix

\section{  Bosonic amplitudes expressed as  matrix permanents}
\label{AppA}

Consider  a $M$-dimensional quantum  unitary network   where $N$ bosons are launched into the  input   modes $|f_1\rangle,\ldots,|f_M\rangle$  and are detected in the output modes $|g_1\rangle,\ldots,|g_M\rangle$. The  network is given  by a $M\times M$-dimensional unitary matrix $U$,  $U^\dag U=I$,  which transforms the input modes into the output modes, 
\be
|f_k\rangle  = \sum_{l=1}^MU_{kl}|g_l\rangle,
\en{A1}
i.e.  between two   orthogonal bases of  a $M$-dimensional single-particle Hilbert space $H$.
The goal is to express the bosonic transition amplitude between two Fock states $|n_1,\ldots,n_M\rangle_f$ and $|m_1,\ldots,m_M\rangle_g$, giving the input and output states of $N$ bosons:
\be
|n_1,\ldots,n_M\rangle_f = \sqrt{\frac{N!}{\prod_{k=1}^{M}n_k!}}|f_{i_1},\ldots,f_{i_N}\rangle,
\en{A2}
\be
|m_1,\ldots,m_M\rangle_g = \sqrt{\frac{N!}{\prod_{k=1}^{M}m_k!}}|g_{j_1},\ldots,g_{j_N}\rangle,
\en{A3}
where $\sum_{k=1}^Mn_k = \sum_{k=1}^Mm_k = N$ and  it is implied that the two sets $\{i_1,\ldots,i_N\}$ and $\{j_1,\ldots,j_N\}$  are composed of repeated mode indices, e.g. 
\be
(i_1,\ldots, i_N) = (\underbrace{1,\ldots,1}_{n_1},\underbrace{2,\dots,2}_{n_2}, \ldots, \underbrace{M,\ldots,M}_{n_M}).
\en{A4}
On the r.h.s.'s of Eqs. (\ref{A2}) and (\ref{A3}) there are unnormalized   symmetric states of $N$ particles in the   tensor product of $N$ single-particle Hilbert spaces $H\otimes H\otimes\ldots\otimes H$. Such a state is given by  a sum over all permutations $\tau$ of   $N$ indices of the single-particle states, divided by the number of all permutations, e.g.
\be
|f_{i_1},\ldots,f_{i_N}\rangle = \frac{1}{N!}\sum_{\tau}|f_{i_{\tau(1)}}\rangle\otimes \ldots \otimes |f_{i_{\tau(N)}}\rangle.
\en{A5}
The bosonic transition amplitude between the Fock states of Eqs. (\ref{A2}) and (\ref{A3})   is given by a double sum  over the two sets of permutations of   indices in  the inner product of  the $f$ and $g$ states, i.e. the indices of   elements of the network matrix $U$. This double sum   is converted to a single one over all permutations of the column indices  by transferring  one of  the two permutations to the co-product indices, i.e.
\begin{eqnarray}
&&{}_g\langle m_1,\ldots,m_M|n_1,\ldots,n_M\rangle_f = \left(\prod_{k=1}^Mm_k!n_k!\right)^{-\frac12}
\nonumber\\
&&\times\frac{1}{N!}\sum_\sigma\sum_\tau\langle g_{j_\sigma(1)}|f_{i_\tau(1)}\rangle\cdot\ldots\cdot\langle g_{j_\sigma(N)}|f_{i_\tau(N)}\rangle
\nonumber\\
&&= \left(\prod_{k=1}^Mm_k!n_k!\right)^{-\frac12}\frac{1}{N!}\sum_\sigma\sum_\tau U_{i_{\tau(1)},j_{\sigma(1)}}\cdot\ldots\cdot U_{i_{\tau(N)},j_{\sigma(N)}}
\nonumber\\
&& = \left(\prod_{k=1}^Mm_k!n_k!\right)^{-\frac12}\sum_{\sigma^\prime} U_{i_1,j_{\sigma^\prime(1)}}\cdot\ldots\cdot U_{i_N,j_{\sigma^\prime(N)}},
\label{A6}\end{eqnarray}
where $\sigma^\prime \equiv \sigma\cdot\tau^{-1}$ runs over all permutations of $N$ elements. One immediately recognizes a matrix permanent on the r.h.s. of Eq. (\ref{A6}), where the matrix consist of repeated rows and columns of the network matrix $U$ (where the order is insignificant due to permutational invariance of  matrix permanent). Therefore, we introduce the notation $U[n_1,\ldots,n_M|m_1,\ldots,m_M]$ for such a matrix and obtain the resulting bosonic  amplitude proportional to permanent of this matrix, as in Eq. (\ref{E1}) of section \ref{sec2} (see also Refs.~\cite{C,TT,Scheel}).

\section{  Approximating the multinomial coefficient}
\label{AppB}

An approximation of the multinomial coefficient, and hence of the Fisher-Yates distribution, can be based on an exact formula for the factorial for $n\ge1$\cite{Mortici}:
\be
n! = \sqrt{2\pi (n+\theta_n)}\left(\frac{n}{e}\right)^n,
\en{B1}
where   $\theta_n$ is  tightly bounded: $1/6<\theta_n<0.177$.  Interestingly, Eq. (\ref{B1})  can be extended to all $n\ge 0$ by carefully defining the limit $0^0 = 1$ and redefining the lower bound to $\theta_0 = \frac{1}{2\pi} <\frac{1}{6}$.
After some   algebraic manipulations, the  multinomial coefficient becomes
 \be
\frac{N!}{\prod_{k=1}^Mn_k!} = \frac{\exp\left(N\mathcal{H}(\{\frac{n_k}{N}\})\right)}{\sqrt{(2\pi N)^{M-1}}}\left(\frac{1+\frac{\theta_N}{N}}{ \prod_{k=1}^M\left[\frac{n_k}{N}+\frac{\theta_{n_k}}{N} \right]}\right)^{\frac12},
\en{B2}
where $\mathcal{H}$ is  the expected   Shannon entropy   function $\mathcal{H}(\{\frac{n_k}{N}\}) \equiv -\sum_{k=1}^M\frac{n_k}{N}\ln\left(\frac{n_k}{N}\right)$. Since $\theta$ in   Eq. (\ref{B2}) is always divided by $N\gg1$,  the simplest approximation is to drop it, thereby making an  error of order  $\mathcal{O}(N^{-1})$ and restricting ourselves to $n_k\ge1$. Assuming the latter, we obtain
\be
\frac{N!}{\prod_{k=1}^Mn_k!} = \frac{\exp\left(N\mathcal{H}(\{\frac{n_k}{N}\})\right)}{\sqrt{(2\pi N)^{M-1}\prod_{k=1}^M\frac{n_k}{N}}}(1 +\mathcal{O}(N^{-1})).
\en{B3}

Finally, an even better  approximation of the multinomial coefficient (for all $n\ge 0$) could be obtained  if  an uniform nonzero $\theta_n$ is selected, for instance  $\theta_n = 1/6$, i.e.  similar as in  Gosper's  approximation of the factorial \cite{Gosper}. Such a formula, though being more complicated,  would then improve the   approximation of the Fisher-Yates distribution of section \ref{sec2}. However, for the sake of simplicity, we do not pursue this approach in the present work.

\section{  Proofs of the determinant identities of section \ref{sec2}}
\label{AppC}

Let us first proof a generalization of Sylvester's determinant identity, i.e.  Eq.  (\ref{E19}) of section \ref{sec2}.
To this goal, one can use the following auxiliary Gaussian integral 
\be
J = \int\limits_{R^n} d^nx e^{-\widetilde{x}Ax}\prod_{j=1}^m \delta\left(\sum_{i=1}^nC_{ji}x_i\right),
\en{C1}
where, for convenience, we replace $C_{j,kl}$ by $C_{ji}$, i.e. the double index by  a single one, and denote $m = 2M-1$ and $n = M^2$.
Here $\delta(y)$ is the Dirac delta function, $\widetilde{x} = (x_1, \ldots, x_n)$, $\mathrm{det}(A)\ne0$, and $\mathrm{rank}(C) = m$.  Our determinant identity (\ref{E19}) follows if we evaluate $J$ by two different methods.  Let us  assume   that the real part in the Hermitian  decomposition of $A$, $A = A_R + iA_I$, is positive: $A_R>0 $. The first method consists of  direct integration over  $n-m$ independent  variables  $y$ extracted   from the variables  $x$. Suppose that the full rank submatrix  $C^{(I)}$ is given by the first $m$ columns of $C$, thus  $C^{(II)}$ is the remaining  columns.  Then $\widetilde{y} = (x_{m+1},\ldots, x_n)$. Resolving the constraints given by the Dirac delta functions in Eq. (\ref{C1}) we express  the dependent variables  as $x^{(I)} = By$, where $B \equiv  - (C^{(I)})^{-1}C^{(II)}$,  thus the whole vector  is as follows
\be
 x = \left[\begin{array}{c} B\\ I \end{array}\right]y.
\en{C2}
The integral in Eq. (\ref{C1}) becomes
\be
J = \frac{1}{|\mathrm{det}(C^{(I)})|}\int \limits_{R^{n-m}}d^{n-m}y \exp\left\{ -\widetilde{y}[\widetilde{B},I]A\left[\begin{array}{c} B\\ I \end{array}\right] y\right\},
\en{C3}
where the real part of the quadratic form in the exponent is positive definite. The Gaussian  integral in Eq. (\ref{C3}) can be easily evaluated and we obtain
\be
J = \sqrt{ \pi^{n-m} }|\mathrm{det}(C^{(I)})|^{-1}\mathrm{det}\left([\widetilde{B},I]A\left[\begin{array}{c} B\\ I \end{array}\right]\right)^{-\frac12}.
\en{C4}
On the other hand, one can also use the Fourier representation of the Dirac delta functions in the integrand of  Eq. (\ref{C1}) and, by interchanging the integration order, evaluate the integral $J$ on the whole $x$ space and then take the inverse Fourier transform. As all  integrals are Gaussian they are easily evaluated. We obtain 
\begin{eqnarray}
&&J = \int\limits_{R^n}d^n x \int\limits_{R^m}\frac{d^m \lambda}{(2\pi)^m} \exp\{-\widetilde{x}Ax + i\widetilde{\lambda}Cx\}
\nonumber\\
&& = \int\limits_{R^m}\frac{d^m \lambda}{(2\pi)^m} \frac{\pi^{n/2}}{ \sqrt{\mathrm{det}(A)} } \exp\{-\frac{1}{4}\widetilde{\lambda}CA^{-1}\widetilde{C}\lambda\}
\nonumber\\
&& = \sqrt{\pi^{n-m}}\left[ \mathrm{det}(A) \mathrm{det}(CA^{-1}\widetilde{C})\right]^{-\frac12}.
\label{C5}
\end{eqnarray}
Comparison of  Eqs. (\ref{C4}) and (\ref{C5})  gives   the determinant identity of Eq.  (\ref{E19})   for nonsingular matrices $A$ with a  positive definite real part. The validity can be extended to  arbitrary nonsingular matrices $A$ by  uniqueness of the analytic continuation in the complex plane, by  noticing  that the r.h.s.'s of Eqs. (\ref{C4}) and (\ref{C5}) are analytic functions of the elements of $A$.

Now, let us verify that all  $(2M-1)\times(2M-1)$-dimensional principal submatrices  of matrix $D$ (\ref{E24}), i.e.  obtained by crossing  out  one row and  one column with the same index, have  equal determinant. This is a consequence of  existence of a  unique  $2M$-dimensional null eigenvector  of $D$, $Dv = 0$,  where $\widetilde{v} =  (1,\ldots,1,-1,\ldots,-1)$ (which easily follows from  the definition (\ref{E24}) and that $\mathrm{rank}(C) = 2M-1$).   Consider  now the  adjoint matrix $\hat{D}$ composed of the minors of $D$ (the adjoint of $A$ is the matrix $\hat{A}$ satisfying $\hat{A}A = \mathrm{det}(A)I$). The principal minors of $D$ are   diagonal elements of $\hat{D}$, thus we have to verify that  all diagonal elements of $\hat{D}$ are equal. We have: $\hat{D}D=0$ which is possible  only if there is a  vector $u$ such that $\hat{D} = u\widetilde{v}$. Since $D$ is a symmetric matrix, such is also $\hat{D}$ and we have $\hat{D} = \alpha v\widetilde{v}$ for some scalar $\alpha$. The latter equality means  $\hat{D}_{jj} = \alpha$ and hence the $(2M-1)\times(2M-1)$-dimensional principal minors of $D$  are all equal. This  allows us to use  any of the  $(2M-1)\times(2M-1)$-dimensional  principal minors of $D$   in the denominator on the r.h.s. in Eq. (\ref{E22}).

Finally, let us simplify  the expression for a $(2M-1)\times(2M-1)$-dimensional principal minor of $D$ (\ref{E24}). To this goal the following determinant  identity  valid for $2\times2$-block matrices  can be used
\begin{eqnarray}
&&\mathrm{det}\left(\begin{array}{c c} A_1 & A_2 \\
A_3 & A_4 \end{array}\right) = \mathrm{det}(A_1) \mathrm{det}(A_4 - A_3A^{-1}_1A_2) \nonumber\\
&& \qquad =\mathrm{det}(A_4)\mathrm{det}(A_1 - A_2A^{-1}_4A_3),
\label{C6}\end{eqnarray}
which is a   generalization of the formula for  determinant of $2\times2$-dimensional matrices.  For the $(2M-1)\times(2M-1)$-dimensional principal minors  of $D$  we  obtain
\begin{eqnarray}
\mathrm{det}(D^\prime)& =&\left[ \prod_{k=1}^M\frac{n_k}{N}\right]\mathrm{det}\left( \Lambda^\prime_2 - \widetilde{p}^\prime \Lambda^{-1}_1p^\prime\right)
\nonumber\\
&=&\left[ \prod_{k=1}^M\frac{m_k}{N}\right]\mathrm{det}\left( \Lambda^\prime_1 -  p^\prime \Lambda^{-1}_2 \widetilde{p}^\prime \right),
\label{C7}\end{eqnarray}
where we have denoted $\Lambda_1 = \mathrm{diag}(\frac{n_1}{N},\ldots,\frac{n_M}{N})$, $\Lambda_2 = \mathrm{diag}(\frac{m_1}{N},\ldots,\frac{m_M}{N})$, whereas the   matrices $\Lambda_{1,2}^\prime$ and $p^\prime$  are those appearing in  the submatrix  $D^\prime$.


\section{  On the computational complexity of  the permanent of a matrix with repeated rows and/or columns }
\label{AppD}

One can reduce the number of operations  in   Ryser's formula \cite{Ryser}  giving the matrix permanent  when the matrix has repeated rows or columns, (see also Appendix B in Ref.~\cite{Thesis}).  Indeed, Ryser's formula uses the inclusion and exclusion principle of Sylvester, it can be cast as
\be
\mathrm{per}(A) = \prod_{i=1}^N \sum_{j=1}^N A_{ij} - \sum_{S_1}\prod_{i=1}^N\sum_{j\in S_1}A_{ij} +\ldots+(-1)^{N-1}\sum_{S_{N-1}}\prod_{i=1}^N\sum_{j\in S_{N-1}}A_{ij},
\en{D1}
where   $S_R \subset \{1,\ldots,N\}$ and has $N-R$ elements (hence, the first term  corresponds to $S_0$). In Eq. (\ref{D1}) the $R$th  term is a sum over  the products of row sums of the  matrices extracted  from $A$ by crossing out $R$ columns. Now let us consider the $N\times N$-dimensional matrix $U[n_1,\ldots,n_M|m_1,\ldots,m_M]$ of section \ref{sec2}. Introducing the notation $U_{k_il_j}$ for its $(i,j)$th element  (where $i$ and $j$ run from 1 to $N$, while  $1\le k_i,l_j\le M$) we get from Eq. (\ref{D1})
\be
\sum_{S_R} \prod_{i=1}^N \sum_{j\in S_R} U_{k_il_j}  =\! \sum_{r_1=0}^{m_1}\!\ldots\!\sum_{r_M=0}^{m_M} \!\!\delta_{\sum_{k=1}^Mr_k,R}
\prod_{k=1}^M\left\{\frac{m_k!}{(m_k-r_k)! r_k!}\left[\sum_{l=1}^M (m_l\!-\!r_l)U_{kl}\right]^{n_k}\right\}
\en{D2}
with  the summation over $\{S_R\}$ being reduced  to that over $\{r_1,\ldots,r_M\}$ with $r_1+\ldots+r_M=R$ ($r_l$ is the number of the $l$th column duplicates  crossed out from $U[n_1,\ldots,n_M|m_1,\ldots,m_M]$).   One can now easily estimate the number of floating point operations (flops) necessary for computing the permanent of $U[n_1,\ldots,n_M|m_1,\ldots,m_M]$. Indeed, computation of the $R$th term by Eq. (\ref{D2})  involves $T_R$ summations over $\{r_1,\ldots,r_M\}$ and $N$ multiplications of a sum comprised of  between 1 and $M$ elements $U_{kl}$. The worst case for the last two operations is thus $MN$ flops. On the other hand, the number $T_R$ can  expressed as 
\be
T_R = \sum_{r_1=0}^{m_1}\!\ldots\!\sum_{r_M=0}^{m_M}  \delta_{\sum_{k=1}^Mr_k,R} = \frac{1}{R!}\frac{d^R}{dz^R}P_N(z)\biggr|_{z=0},
\en{D3}
where $P_N(z) = \prod_{k=1}^M(1+z+\ldots+z^{m_k})$. Thus  $T_R$ is the $R$th term in the Taylor expansion of $P_N(z)$ about $z=0$ and with $\Delta z = 1$. While   $T_R$ seems to be given by a complicated   dependence on  $R$ and $\{m_l\}$, their sum, i.e. the total number of flops in the summations over all sets $\{S_R\}$, has a simple expression. Indeed, using the fact that the Taylor expansion for $P_N(z)$ has only $N+1$ terms, we get
\be
\sum_{R=0}^{N-1} T_R = P_N(1) - 1 = \prod_{k=1}^M(m_k+1) -1. 
\en{D4}
Using this result and the previous estimates on the number of flops in the product and summation inside each sum over $\{r_1,\ldots,r_M\}$, as in   Eq. (\ref{D2}), we get that the number of necessary flops   $\mathcal{F}$ in Ryser's formula can be reduced to  a value satisfying 
\be
N\left[ \prod_{k=1}^M(m_k+1) -1\right] < \mathcal{F} < MN \left[\prod_{k=1}^M(m_k+1) -1\right]
\en{D5}
(note that by setting $M=N$ and $m_k = 1$  we get the well-known upper estimate on the number of flops necessary  for computing the permanent of an arbitrary matrix by  Ryser's formula:   $\mathcal{N}=\mathcal{O}( N^22^N)$). Let  us now analyze the worst case which is obtained by uniformly distributing the bosons over the modes, i.e. when $m_k =   {N}/{M}$ (this maximizes the product in Eq. (\ref{D5})). We obtain   the upper estimate on the number of flops (assuming $N\gg M$, and $M$ fixed)
\be
\mathcal{F}  = \mathcal{O}\left( {N^{M+1}}\right).
\en{D6}
This result   conforms with the general guess,  stated in the Introduction,  on  the number of necessary flops for computing the permanent of a $N\times N$-dimensional matrix of rank $M$.  

Finally we note that complete characterization of a   network for  a given input $\{n_1,\ldots,n_M\}$ is given by probabilities of  all  possible distributions 
$\{m_1,\ldots,m_M\}$ of particles  in    the output modes.  Hence, to characterize   $N$ bosons on   a $M$-mode    network with $N\gg M$ and $M$ fixed  one has to compute 
\be
\mathcal{N} = \sum_{m_1=0}^N\ldots \sum_{m_M=0}^N \delta_{\sum_{k=1}^Mm_k,N } =  \frac{(M+N-1)!}{(M-1)!N!} = \mathcal{O}(N^{M-1})
\en{D7}
permanents. Thus the total  number of necessary  flops to characterize   $N$ bosons on   a $M$-mode    network with $N\gg M$ and $M$ fixed  is $\mathcal{F}_{Total} =  \mathcal{O}(N^{2M})$. 


\end{document}